\documentclass[journal]{IEEEtran}  
\IEEEoverridecommandlockouts                              

\usepackage[T1]{fontenc}
\usepackage[utf8]{inputenc}
\usepackage{cite}
\usepackage{amsmath,amssymb,amsfonts}
\usepackage{algorithmic}
\usepackage{graphicx}
\usepackage{subcaption}
\usepackage{textcomp}
\usepackage{xcolor}
\usepackage{booktabs}
\usepackage{multirow}
\usepackage{array}
\usepackage{url}
\usepackage{hyperref}

\hypersetup{
    colorlinks=true,
    linkcolor=black,
    citecolor=black,
    urlcolor=blue
}

\title{Lightweight Adaptation of EEG Foundation Models for Stroke Motor Imagery Decoding: Domain Shift and Subject-Level Robustness}

\author{Anh T. Nguyen, \textit{Member, IEEE}$^{1\dagger}$, Zihua Sun$^{2\dagger}$, and Michelle J. Johnson, \textit{Member, IEEE}$^{3}$
\thanks{*This work was supported by the American Heart Association (AHA) Predoctoral Fellowship under Award Number 25PRE1372607 and the Vingroup Scholarship Program for Master’s and Ph.D. Degrees Overseas Study.}
\thanks{$^{1}$Anh T. Nguyen is a PhD Candidate with the School of Engineering and Applied Science, Department of Bioengineering, University of Pennsylvania, Philadelphia, PA, USA {\tt\small tuna28ng@engineering.upenn.edu}}
\thanks{$^{2}$Zihua Sun is a M.S. student with the School of Engineering and Applied Science, Department of Electrical and Systems Engineering, University of Pennsylvania, Philadelphia, PA, USA
    {\tt\small szh16@engineering.upenn.edu}}%
\thanks{$^{3}$Dr. Michelle J. Johnson is a Professor with the Department of Physical Medicine and Rehabilitation, Department of Bioengineering, and Department of Mechanical Engineering and Applied Mechanics, University of Pennsylvania, Philadelphia, PA, USA {\tt\small johnmic@pennmedicine.upenn.edu}}
\thanks{$^{\dagger}$ These authors contributed equally to this work.}%
}

\begin{document}
\maketitle
\thispagestyle{empty}
\pagestyle{empty}

\begin{abstract}
Motor imagery (MI) electroencephalography (EEG) decoding could support post-stroke rehabilitation, but models developed on healthy cohorts may not transfer reliably to pathological EEG. We evaluated whether Low-Rank Adaptation (LoRA) can efficiently adapt three pretrained EEG foundation models (i.e., LaBraM-base, REVE-base, and REVE-large) for binary left- versus right-hand MI decoding. Frozen-backbone head-only baselines and LoRA adaptation were evaluated using subject-wise five-fold cross-validation on the PhysioNet EEG Motor Movement/Imagery Dataset and a binary subset of the UET175 dataset comprising 30 stroke participants. 
On EEGMMIDB, LoRA increased accuracy to 0.822 for LaBraM-base and 0.957 for REVE-base. On UET175, all head-only models performed near chance. With LoRA, LaBraM-base remained near chance (0.499$\pm$0.009), whereas REVE-base reached 0.847$\pm$0.194 and outperformed REVE-large (0.806$\pm$0.178), indicating that increased model capacity alone did not improve stroke-domain adaptation. 
The strongest stroke configuration, REVE-base LoRA, was further evaluated using within-cohort leave-one-subject-out cross-validation (LOOCV), showing 0.952 mean accuracy, but subject-wise accuracy ranged from 0.586 to 1.000, revealing a small low-performing tail. 
Zero-shot transfer from EEGMMIDB to UET175 remained near chance (0.464$\pm$0.072). These findings show that healthy-benchmark performance does not ensure transfer to stroke EEG. 
Translation of EEG foundation models to pseudo-online or real-time rehabilitation BCIs should therefore include target-domain adaptation and subject-level assessment of temporal informativeness, spatial sensitivity, and physiological discriminability.
\end{abstract}


\section{Introduction}

Electroencephalography (EEG)-based motor imagery (MI) brain-computer interfaces (BCIs) have been widely studied for assistive control and neurorehabilitation because they provide a non-invasive pathway for decoding movement intention from brain activity \cite{padfield2019mi_review}. In stroke rehabilitation, MI is particularly important because it can engage sensorimotor processes even when overt movement is limited, and MI-BCI systems have shown promise for improving motor recovery and supporting active rehabilitation training \cite{mulder2007mi_rehab,liao2023mi_bci_rehab,wang2024bci_ul_rehab}. At the same time, practical deployment remains challenging, as clinical EEG is highly variable across patients, signal quality is inconsistent, and decoding performance often drops when models are transferred from healthy research datasets to pathological populations \cite{khan2020stroke_bci_review,liao2023mi_bci_rehab}.

Conventional MI decoding pipelines have largely relied on handcrafted feature extraction, compact convolutional networks, or dataset-specific classifiers \cite{padfield2019mi_review,lawhern2018eegnet}. While such approaches can perform well in controlled settings, they typically require careful task-specific design and often struggle with cross-subject generalization and deployment robustness. This limitation is especially relevant in stroke applications, where lesion location, impairment severity, and subject-specific neural reorganization can introduce substantial inter-subject heterogeneity \cite{khan2020stroke_bci_review,liao2023mi_bci_rehab}. Recent work has also emphasized that offline accuracy alone may not adequately reflect realistic BCI usability, motivating more deployment-oriented evaluation strategies such as pseudo-online or adaptive analyses \cite{carrara2024pseudo_online,acqualagna2016coadaptive,choi2024real_time_robot}.


EEG foundation models have recently emerged as a promising alternative to conventional task-specific MI decoding pipelines. By pretraining on large and diverse EEG corpora, these models aim to learn reusable neural representations that can reduce dependence on handcrafted features and improve transfer to downstream decoding tasks \cite{kuruppuEEGFoundationModels2026}. 
At the same time, parameter-efficient fine-tuning methods provide a practical way to adapt the pretrained EEG foundation models with limited data and compute. However, it remains unclear whether these advantages persist in clinically relevant settings such as post-stroke MI decoding, where healthy-to-stroke transfer introduces substantial domain shift in physiology, signal quality, and inter-subject variability. This gap motivates our study of whether pretrained EEG foundation models can be lightly adapted for stroke MI decoding, and whether their performance remains robust beyond cohort-average accuracy.

In this paper, we study lightweight adaptation of EEG foundation models for stroke MI decoding. We first benchmark publicly available pretrained models \cite{jiang2024labram,ouahidi2025reve} in a head-only setting on two MI datasets comprising healthy participants and stroke survivors \cite{eegmmidb_physionet, uet175_dataset}, then evaluate the main fine-tuning adaptation analysis on these EEG foundation models. 
Beyond this primary comparison, we perform a set of robustness-oriented secondary analyses on the strongest practical configuration, including zero-shot healthy-to-stroke transfer, preprocessing ablation, subject-level analysis, and channel and time-window ablations. These follow-up analyses are used to test whether apparent cohort-level gains remain reliable under severe healthy-to-stroke domain shift and across individual stroke subjects.


\section{Related Work}

\subsection{Motor Imagery EEG Decoding and Stroke Rehabilitation}

Motor imagery (MI) EEG decoding has been widely studied for brain-computer interfaces (BCIs), with applications ranging from communication and assistive control to neurorehabilitation \cite{padfield2019mi_review}. In stroke rehabilitation, MI is particularly relevant because it can engage sensorimotor networks even when physical movement is impaired, and MI-based BCI training has shown potential for improving upper-limb recovery and promoting neuroplasticity \cite{mulder2007mi_rehab,khan2020stroke_bci_review,liao2023mi_bci_rehab,wang2024bci_ul_rehab}. However, decoding stroke EEG remains substantially more difficult than decoding healthy EEG due to pathological variability, lesion-dependent reorganization, lower signal consistency, and strong inter-subject heterogeneity \cite{khan2020stroke_bci_review,liao2023mi_bci_rehab}.

Traditional MI decoding methods have typically relied on handcrafted features such as band power, common spatial patterns, or connectivity features, followed by shallow classifiers \cite{padfield2019mi_review}. Deep learning approaches such as EEGNet have also improved end-to-end decoding performance while keeping model size compact \cite{lawhern2018eegnet}. Nevertheless, many existing approaches remain highly task-specific and dataset-dependent, and their performance often degrades in cross-subject or clinically heterogeneous settings. This limitation is particularly important for rehabilitation BCIs, where robustness across patients is essential for practical use.

\subsection{EEG Foundation Models}


Recent work on EEG foundation models extends earlier self-supervised EEG representation learning toward larger-scale pretraining and broader downstream transfer \cite{kuruppuEEGFoundationModels2026}. Early studies such as BENDR showed that transformer-based self-supervised learning could extract useful EEG representations from large unlabeled corpora \cite{bendr2021}. More recent foundation models explicitly target cross-dataset generalization. Jiang et al. introduced LaBraM, a large-scale pretrained EEG model based on masked modeling over channel-wise temporal patches, aiming to learn generic BCI representations from diverse EEG datasets \cite{jiang2024labram}. REVE, developed by Ouahidi et al., later emphasized a different design priority: handling heterogeneous EEG setups through flexible spatio-temporal embeddings that support varied electrode layouts and signal lengths \cite{ouahidi2025reve}. Together, these models suggest that large-scale pretraining may provide more reusable EEG representations than conventional task-specific pipelines \cite{kuruppuEEGFoundationModels2026}.

Recent benchmarking efforts indicate that EEG foundation modeling is still an open and evolving area rather than a settled solution. Comparative studies such as EEG-FM-Bench and NeuralBench report that results remain sensitive to evaluation protocol, task organization, and fine-tuning strategy, and that larger pretrained models do not uniformly outperform smaller or task-specific alternatives \cite{xiongEEGFMBenchComprehensiveBenchmark2025,banvilleNeuralBenchUnifyingFramework2026}. This literature is directly relevant to our study. We focus on LaBraM and REVE because they are representative, publicly available models that embody two distinct approaches to EEG foundation modeling: LaBraM as a masked-pretraining-based model for generic EEG representation learning, and REVE as a heterogeneity-aware architecture explicitly designed to support transfer across different EEG setups \cite{jiang2024labram,ouahidi2025reve}.

However, most existing evaluations still emphasize standard research datasets or non-clinical downstream tasks \cite{xiongEEGFMBenchComprehensiveBenchmark2025,banvilleNeuralBenchUnifyingFramework2026}. Whether such pretrained representations remain effective under severe healthy-to-stroke domain shift, and whether their benefits persist at the subject level in stroke MI decoding, remains largely untested. 
Rather than treating performance on healthy EEG benchmarks as a proxy for clinical utility, we test whether representative EEG foundation models can be practically adapted for binary stroke MI decoding, and whether their benefits remain robust at the subject level in a rehabilitation-relevant setting.

\subsection{Parameter-Efficient Fine-Tuning for EEG}

Parameter-efficient fine-tuning has become an important strategy for adapting large pretrained models to downstream tasks without the cost of full fine-tuning. LoRA is one of the most widely used lightweight fine-tuning methods because it freezes the pretrained backbone and learns low-rank updates with only a small number of trainable parameters \cite{hu2021lora}. This makes it particularly attractive for EEG applications, where downstream datasets are often small and computational resources may be limited.

Recent studies have started to extend parameter-efficient fine-tuning ideas to EEG. Suzumura et al. introduced GraphAdapter, a graph-based adapter for EEG foundation models by explicitly leveraging spatial structure \cite{suzumura2024graphadapter}. Lotey et al. proposed an ensemble of weight-decomposed low-rank adaptation method called EDoRA for mental imagery EEG tasks \cite{lotey2024edora}. Wang et al. explored Task-Conditioned Prompt Learning, a prompt-based adaptation method for few-shot cross-subject MI decoding \cite{wang2025tcpl}. These works suggest that lightweight adaptation can be effective in EEG settings. However, they do not directly address a clinically important scenario, which is characterized by adapting pretrained EEG foundation models from healthy MI benchmarks to stroke MI decoding under pathological domain shift.

\section{Methods}



\subsection{Datasets}

\subsubsection{EEGMMIDB Healthy Reference Dataset}

We use the publicly available EEG Motor Movement/Imagery Dataset (EEGMMIDB) from PhysioNet as the healthy-subject reference dataset for binary left-hand versus right-hand MI decoding \cite{eegmmidb_physionet}. 
This dataset contains 64-channel EEG recordings acquired using the BCI2000 system from 109 healthy volunteers during both executed and imagined motor tasks. EEG signals were recorded using the International 10-10 electrode placement system (Fig. \ref{fig:easycap_all64}), at a sampling rate of 160 Hz.
The relevant MI runs required participants to imagine repeatedly opening and closing either the left or right fist in response to visual cues.
In this study, we use only the MI runs corresponding to unilateral fist movements (runs R04, R08, and R12), where subjects imagined either left-hand or right-hand fist movements. Specifically, event label T1 corresponds to left-fist imagery and T2 corresponds to right-fist imagery in these runs.

\subsubsection{UET175 Stroke Dataset}

We use the public UET175 dataset as the post-stroke reference dataset for MI decoding in stroke patients \cite{uet175_dataset}. 
The dataset contains EEG recordings collected from 30 post-stroke patients using the Emotiv EPOC Flex system at a sampling rate of 128 Hz.
Although the acquisition setup supported 32 electrodes, the dataset specifically selected 22 channels located near the sensorimotor cortex: FZ, FC3, FC1, FCZ, FC2, FC4, C5, C3, C1, CZ, C2, C4, C6, CP3, CP1, CPZ, CP2, CP4, P1, PZ, P2, POZ (Fig. \ref{fig:easycap_uet22}).
During the recording sessions, participants performed four MI tasks corresponding to four imagined movements, including lifting the left arm, lifting the right arm, lifting the left leg, and lifting the right leg.

\subsection{Preprocessing Pipelines}
Preprocessing was performed using MATLAB (The MathWorks, Inc., Natick, MA, USA) and the EEGLAB toolbox \cite{delorme2004eeglab}.

Raw EEG datasets undergo a standard preprocessing pipeline.
For EEGMMIDB, raw EEG recordings are band-pass filtered from 0.1 to 75\,Hz and notch filtered at 60\,Hz to suppress power-line interference. Signals are resampled to 200\,Hz when needed, restricted to a fixed 64-channel montage, reordered to match the model-required channel order, epoched into 4.0\,s trials, and normalized by dividing microvolt values by 100.


For the public UET175 recordings, the recorded signals had already undergone the EEG device's built-in 5th-order Sinc filtering and 0.2--45\,Hz band-pass filtering \cite{uet175_dataset}. In addition, in the dataset, the event timestamp files released with each run identify accepted motor imagery trials after noisy trials and trials without significant event-related desynchronization were excluded by the dataset authors. Using these timestamp files, we extracted the accepted trials and assigned labels according to the original trial indices. We then retained only the left- and right-hand motor imagery classes, resampled the signals from 128\,Hz to 200\,Hz, and normalized the microvolt values by dividing by 100.
The resulting study-specific analytic dataset is referred to hereafter as the processed binary UET175 subset, or the UET175 subset for brevity. This terminology distinguishes it from the full public UET175 dataset, which contains four MI classes.

\subsection{Foundation Models}

We evaluate three publicly available pretrained EEG foundation models:

\begin{itemize}
\item \textbf{LaBraM-base} \cite{jiang2024labram}
\item \textbf{REVE-base} \cite{ouahidi2025reve}
\item \textbf{REVE-large} \cite{ouahidi2025reve}
\end{itemize}


Although LaBraM and REVE both use patch-based Transformer encoders to learn representations from multichannel EEG, they differ substantially in their spatial parameterization and self-supervised pretraining objectives. LaBraM maps non-overlapping channel-wise waveform patches through a temporal convolutional encoder, adds learned spatial embeddings indexed by electrode identity and learned temporal embeddings indexed by patch position, and is pretrained to predict discrete spectrum-informed neural tokens for masked patches. In contrast, REVE linearly projects overlapping waveform patches and derives their spatio-temporal positions from the electrode's three-dimensional coordinates and temporal patch indices using a four-dimensional Fourier encoding. REVE is pretrained using a masked-autoencoding objective that reconstructs the raw waveforms of masked patches. Thus, LaBraM combines learned electrode-identity embeddings with discrete neural-token prediction, whereas REVE combines geometry-derived positional representations with raw-signal reconstruction. This architectural distinction is particularly relevant to the present study because EEGMMIDB and UET175 differ substantially in electrode montage, acquisition hardware, task paradigm, and participant population.

Let $\mathbf{X}\in\mathbb{R}^{C\times T}$ denote an EEG trial with $C$ channels and $T$ temporal samples. After resampling, each trial in our experiments contains $T=800$ samples at 200\,Hz. Both models represent each channel as a sequence of temporal waveform patches, but they use different patching operations.

LaBraM partitions each channel into non-overlapping patches of length $w=200$ samples, resulting in
$P_{\mathrm{LaBraM}}=\lfloor T/w\rfloor=4$ patches per channel. REVE uses the same patch length but introduces an overlap of $o=20$ samples, corresponding to a stride of $w-o=180$ samples and
$P_{\mathrm{REVE}}=\lfloor (T-w)/(w-o)\rfloor+1=4$ complete patches per channel. Thus, both models produce $C\times4$ channel--time tokens for the input durations used in this study, although they differ in how the waveform patches are embedded, positionally encoded, and pretrained, as described below.

For LaBraM, each waveform patch $\mathbf{x}_{c,k}\in\mathbb{R}^{w}$ from channel $c$ and temporal index $k$ is mapped to a $d_{\mathrm{L}}$-dimensional representation by a temporal convolutional encoder $f_{\mathrm{temp}}(\cdot)$. LaBraM-base uses an embedding dimension of $d_{\mathrm{L}}=200$ and a 12-layer Transformer encoder with 10 attention heads \cite{jiang2024labram}. Before entering the Transformer, each patch embedding is augmented with learned spatial and temporal embeddings:

\begin{equation}
\mathbf{u}_{c,k}
=
f_{\mathrm{temp}}(\mathbf{x}_{c,k})
+
\mathbf{s}_{c}
+
\boldsymbol{\tau}_{k},
\label{eq:labram_embedding}
\end{equation}

where $\mathbf{s}_{c}$ is a learned spatial embedding indexed by electrode identity and $\boldsymbol{\tau}_{k}$ is a learned embedding indexed by temporal patch position. The resulting channel--time tokens are processed jointly by the Transformer, allowing self-attention to model spatial dependencies across electrodes and temporal dependencies across patches.

LaBraM is pretrained using vector-quantized neural spectrum prediction. A separately trained neural tokenizer first converts each continuous EEG patch into a discrete target token. Given a tokenizer representation $\mathbf{p}_{i}$ and a codebook $\mathcal{V}=\{\mathbf{v}_{j}\}_{j=1}^{K}$, the target token is selected as the nearest normalized codebook entry:

\begin{equation}
q_i
=
\operatorname*{arg\,min}_{j\in\{1,\ldots,K\}}
\left\|
\frac{\mathbf{p}_{i}}{\|\mathbf{p}_{i}\|_2}
-
\frac{\mathbf{v}_{j}}{\|\mathbf{v}_{j}\|_2}
\right\|_2.
\label{eq:labram_tokenization}
\end{equation}

The tokenizer is trained to reconstruct normalized Fourier amplitude and phase representations of the original waveform patches. Consequently, its discrete codebook targets contain spectrum-related information rather than representing waveform similarity alone.

During masked EEG modeling, a subset of channel--time patches $\mathcal{M}$ is replaced by learned mask tokens. The Transformer is trained to predict the tokenizer-assigned target $q_i$ for each masked patch:

\begin{equation}
\mathcal{L}_{\mathcal{M}}
=
-
\sum_{i\in\mathcal{M}}
\log
p_{\theta}
\left(
q_i
\mid
\mathbf{X}^{\mathcal{M}}
\right).
\end{equation}

LaBraM additionally applies the complementary mask $\widetilde{\mathcal{M}}$ in a symmetric masked-modeling strategy. Its complete pretraining objective can therefore be written as

\begin{equation}
\mathcal{L}_{\mathrm{LaBraM}}
=
\mathcal{L}_{\mathcal{M}}
+
\mathcal{L}_{\widetilde{\mathcal{M}}}.
\label{eq:labram_pretraining}
\end{equation}

This symmetric objective enables the model to reuse the same tokenizer targets under complementary masking patterns while learning contextual representations of the EEG patches \cite{jiang2024labram}.

After pretraining, the neural tokenizer and masked-token prediction head are not required for downstream classification. In our implementation, the final Transformer representations of the channel--time patches are mean-pooled and normalized to obtain a trial-level feature vector, which is subsequently passed to the binary MI classification head.

For REVE, each waveform patch $\mathbf{x}_{c,k}\in\mathbb{R}^{w}$ is mapped directly into the model embedding space through a linear projection:

\begin{equation}
\mathbf{e}_{c,k}
=
\mathbf{W}_{p}\mathbf{x}_{c,k}
+
\mathbf{b}_{p}.
\label{eq:reve_patch_embedding}
\end{equation}

REVE-base and REVE-large both use 22-layer Transformer encoders. REVE-base has an embedding dimension of 512 with 8 attention heads, whereas REVE-large increases the embedding dimension to 1,216 with 19 attention heads \cite{ouahidi2025reve}.

Unlike LaBraM's learned electrode-identity embeddings, REVE derives the position of each patch from the three-dimensional coordinate $\mathbf{r}_{c}\in\mathbb{R}^{3}$ of electrode $c$ and the temporal patch index $k$. During pretraining, Gaussian noise is added to the electrode coordinates as a spatial augmentation intended to improve robustness to variability in electrode placement. The resulting four-dimensional spatio-temporal coordinate is

\begin{equation}
\mathbf{q}_{c,k}
=
\left[
(\mathbf{r}_{c}+\boldsymbol{\epsilon}_{c})^{\mathsf T},
\,s_t k
\right]^{\mathsf T}
\in\mathbb{R}^{4},
\qquad
\boldsymbol{\epsilon}_{c}
\sim
\mathcal{N}(\mathbf{0},\sigma^{2}\mathbf{I}),
\label{eq:reve_coordinates}
\end{equation}

where $s_t$ scales the temporal coordinate to a range comparable with the spatial coordinates.

The coordinate $\mathbf{q}_{c,k}$ is mapped through a four-dimensional Fourier basis. For frequency vectors $\boldsymbol{\omega}\in\Omega$, this basis contains sine and cosine components of the form
$\sin(2\pi\boldsymbol{\omega}^{\mathsf T}\mathbf{q}_{c,k})$
and
$\cos(2\pi\boldsymbol{\omega}^{\mathsf T}\mathbf{q}_{c,k})$.
The fixed Fourier features are combined with a learnable nonlinear projection of the same coordinates:

\begin{equation}
\mathbf{r}^{\,\mathrm{pos}}_{c,k}
=
\operatorname{LN}
\left[
\boldsymbol{\Phi}_{\Omega}(\mathbf{q}_{c,k})
+
\operatorname{LN}
\left(
\operatorname{GELU}
\left(
\mathbf{W}_{\mathrm{lin}}\mathbf{q}_{c,k}
\right)
\right)
\right],
\label{eq:reve_position}
\end{equation}

where $\boldsymbol{\Phi}_{\Omega}(\cdot)$ denotes the Fourier feature mapping. The positional representation $\mathbf{r}^{\,\mathrm{pos}}_{c,k}$ is added to the waveform embedding $\mathbf{e}_{c,k}$ before the tokens are processed by the Transformer. Because this positional encoding is computed from electrode geometry and patch time rather than retrieved from a fixed electrode-identity table, REVE can construct representations for different electrode layouts and input durations, provided that the corresponding electrode coordinates are available \cite{ouahidi2025reve}.

REVE is pretrained using a modified masked-autoencoding objective. During pretraining, contiguous groups of tokens are masked jointly across the spatial and temporal dimensions. Only the visible tokens are passed through the Transformer encoder. A lightweight decoder then combines the encoded visible representations with learned mask tokens and reconstructs the raw EEG waveforms of the masked patches. The primary reconstruction loss is

\begin{equation}
\mathcal{L}_{\mathrm{primary}}
=
\frac{1}{|\mathcal{M}|}
\sum_{i\in\mathcal{M}}
\left\|
\widehat{\mathbf{x}}_{i}
-
\mathbf{x}_{i}
\right\|_{1}.
\end{equation}

In addition to the primary reconstruction task, REVE uses a secondary task that reconstructs the masked patches from a compact global representation obtained through learned-query attention pooling across encoder features. The complete pretraining objective is

\begin{equation}
\mathcal{L}_{\mathrm{REVE}}
=
\mathcal{L}_{\mathrm{primary}}
+
\lambda
\mathcal{L}_{\mathrm{global}},
\label{eq:reve_pretraining}
\end{equation}

where both terms use an $L_1$ waveform-reconstruction loss. The global objective encourages the encoder to retain information that can be summarized into a trial-level representation rather than relying exclusively on local patch features \cite{ouahidi2025reve}.

After pretraining, the reconstruction decoder is discarded and all unmasked channel--time patches are processed by the encoder. In our downstream implementation, the final token representations
$\{\mathbf{h}_{i}\}_{i=1}^{N}$
are summarized using the learned-query attention-pooling operation provided by REVE \cite{ouahidi2025reve}. Given a learned query vector $\mathbf{a}$,

\begin{equation}
\begin{aligned}
\alpha_i
&=
\frac{
\exp\left(
\mathbf{a}^{\mathsf T}\mathbf{h}_{i}/\sqrt{d}
\right)
}{
\sum_{j=1}^{N}
\exp\left(
\mathbf{a}^{\mathsf T}\mathbf{h}_{j}/\sqrt{d}
\right)
},\\
\mathbf{z}_{\mathrm{REVE}}
&=
\sum_{i=1}^{N}
\alpha_i\mathbf{h}_{i},
\end{aligned}
\label{eq:reve_attention_pooling}
\end{equation}

where $d$ is the embedding dimension. The pooled representation $\mathbf{z}_{\mathrm{REVE}}$ is subsequently passed to the binary MI classification head. This learned-query pooling differs from the mean pooling used in our LaBraM implementation by allowing the contribution of each channel--time token to be weighted adaptively.

The comparison between LaBraM and REVE therefore concerns different representational inductive biases rather than a simple distinction between supporting and not supporting variable EEG configurations. LaBraM can process different channel subsets and signal durations, but its spatial representations are selected from a learned electrode-identity embedding vocabulary. REVE instead generates positional representations directly from electrode coordinates and temporal patch indices. The two models also differ in their self-supervised targets and downstream aggregation: LaBraM predicts discrete spectrum-informed tokens and uses mean-pooled downstream features, whereas REVE reconstructs masked raw waveforms and uses learned-query attention pooling \cite{jiang2024labram,ouahidi2025reve}.

These differences are directly relevant to the present study. EEGMMIDB and UET175 use different electrode montages, acquisition systems, and original sampling rates, and they also differ in task paradigm and participant population. Resampling both datasets to 200\,Hz standardizes the temporal sampling frequency before model input, but it does not remove differences in spatial configuration, recording hardware, motor-imagery task, or healthy-versus-stroke neurophysiology. Our comparison therefore evaluates whether the distinct pretrained representations of LaBraM and REVE remain usable and adaptable under this combined domain shift.

REVE-large was included as a high-capacity reference and model-capacity control. Because REVE-large is substantially larger than REVE-base, the primary lightweight comparison focuses on LaBraM-base and REVE-base, which represent two distinct foundation-model designs while remaining more computationally practical. REVE-large was additionally evaluated with LoRA under the same adaptation strategy to test whether increasing backbone capacity alone improves downstream adaptation under stroke-domain shift.

\subsection{Head-Only Baseline and LoRA Adaptation}

\subsubsection{Head-Only Baseline}
All three models are first evaluated in a frozen-backbone head-only setting, in which the pretrained encoder parameters remain fixed and only a newly added task-specific classification head is optimized. This setting evaluates how much MI-relevant information can be extracted from the pretrained representation without modifying the backbone. 

\subsubsection{LoRA Fine-Tuning}
For lightweight adaptation, we use Low-Rank Adaptation (LoRA), which freezes the pretrained backbone and inserts trainable low-rank updates into selected linear projections. In all LoRA experiments, we use rank $r=8$, scaling factor $\alpha=16$, and no dropout.

For LaBraM-base, LoRA is applied to every linear layer named \texttt{qkv} in the transformer attention blocks, replacing 12 \texttt{qkv} modules. For REVE-base and REVE-large, LoRA is applied to the attention projection modules \texttt{to\_qkv} and \texttt{to\_out} in each attention block. This results in 77,202 trainable parameters out of 5,901,938 total parameters for LaBraM-base and 540,672 trainable parameters out of 69,730,304 total parameters for REVE-base.

REVE-large LoRA is included as a capacity-control experiment to test whether increasing backbone size improves lightweight adaptation.

\subsection{Training and Evaluation Protocol}

Unless otherwise stated, experiments are trained with seed 42, batch size 64, 50 epochs, learning rate $5\times10^{-4}$, weight decay 0.05, and 5 warmup epochs. Training is performed in PyTorch 2.8.0+cu128 on an NVIDIA RTX 4080 Laptop GPU.

For both the EEGMMIDB and UET175-subset experiments, we use 5-fold subject-wise cross-validation. Classification performance is reported using accuracy, balanced accuracy, Cohen's $\kappa$, and weighted F1-score.



\subsection{Secondary Analyses}

The secondary analyses are designed as follow-up evaluations of the strongest setting identified in the primary analysis. Unless a given secondary analysis explicitly compares multiple primary-analysis settings, we use the best-performing model-and-adaptation mode as the main configuration for these follow-up experiments. This design keeps the secondary section focused on robustness, preprocessing sensitivity, and failure-case interpretation of the most practically relevant stroke-decoding setting rather than introducing a second stage of model selection.

\subsubsection{Artifact Subspace Reconstruction-Based Preprocessing Ablation}

In addition to the standard pipeline, we evaluate a preprocessing pipeline variant that uses Artifact Subspace Reconstruction (ASR) algorithm as an ablation. ASR is a data-driven artifact removal method that identifies high-variance artifact subspaces and reconstructs contaminated signal components in EEG recordings \cite{mullen2015asr}; in this study, it was implemented using the EEGLAB function \texttt{clean\_rawdata}. 
In the ASR-based preprocessing pipeline, ASR is used to automatically detect and reject bad channels (e.g., flatline and excessively noisy channels), as well as to repair or remove bad epochs.
For ASR-preprocessed data, the input is normalized by dividing by 40 rather than 100 to account for the altered signal scale after preprocessing.


ASR was evaluated for a targeted subset of configurations rather than for every model-and-mode combination. We applied ASR to the EEGMMIDB head-only baselines and to the best-performing model-and-adaptation setting identified in the primary analysis, using these experiments to test whether stronger artifact removal changes performance in the baseline frozen-feature paradigm and in the main adapted setting on healthy and stroke EEG. Because this was a secondary preprocessing ablation rather than an exhaustive re-evaluation of the full experimental grid, ASR conclusions are restricted to the evaluated configurations.

\subsubsection{Subject-level analysis}
For subject-level robustness analysis on the UET175 dataset, we perform leave-one-subject-out cross-validation (LOOCV) with 30 folds. In each fold, one subject is held out for testing, two subjects are used for validation, and the remaining 27 subjects are used for training. This protocol is used to quantify subject-wise generalization beyond cohort-average performance and to identify individual low-performing cases within the stroke cohort.

The LOOCV results also serve as the basis for follow-up subject-level interpretation. In particular, subjects with persistently low decoding performance are examined in follow-up analyses to assess whether failure can be explained by gross clinical characteristics alone or instead failure reflects more heterogeneous physiological or signal-structure factors.
To support this interpretation, we perform a targeted clinical matching analysis for the lowest-performing subjects identified by LOOCV. Each low-performing case is compared with a higher-performing subject from the same cohort selected for broadly similar available clinical characteristics, including lesion location and reported severity measures such as National Institutes of Health Stroke Scale (NIHSS), Modified Rankin Scale (mRS), and muscle strength when available. This comparison is not intended as a formal matched-cohort analysis, but as a qualitative follow-up used to assess whether large decoding differences can persist despite broadly similar clinical presentation.

\subsubsection{Channel Ablation}

\begin{figure}[ht]
    \centering

    \begin{subfigure}[t]{0.485\linewidth}
        \centering
        \includegraphics[width=\linewidth]{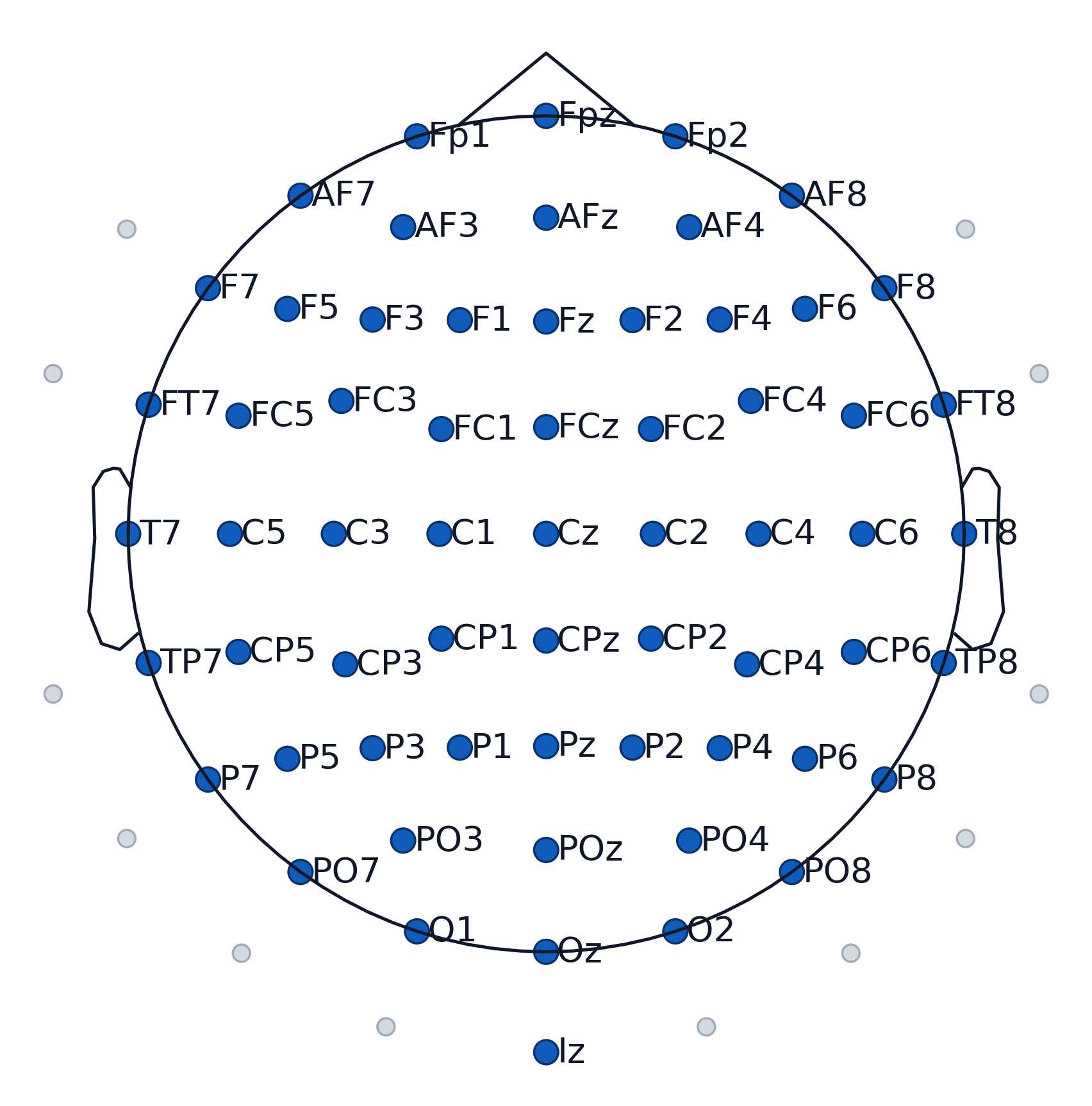}
        \caption{64-channel montage used for EEGMMIDB}
        \label{fig:easycap_all64}
    \end{subfigure}
    \hfill
    \begin{subfigure}[t]{0.485\linewidth}
        \centering
        \includegraphics[width=\linewidth]{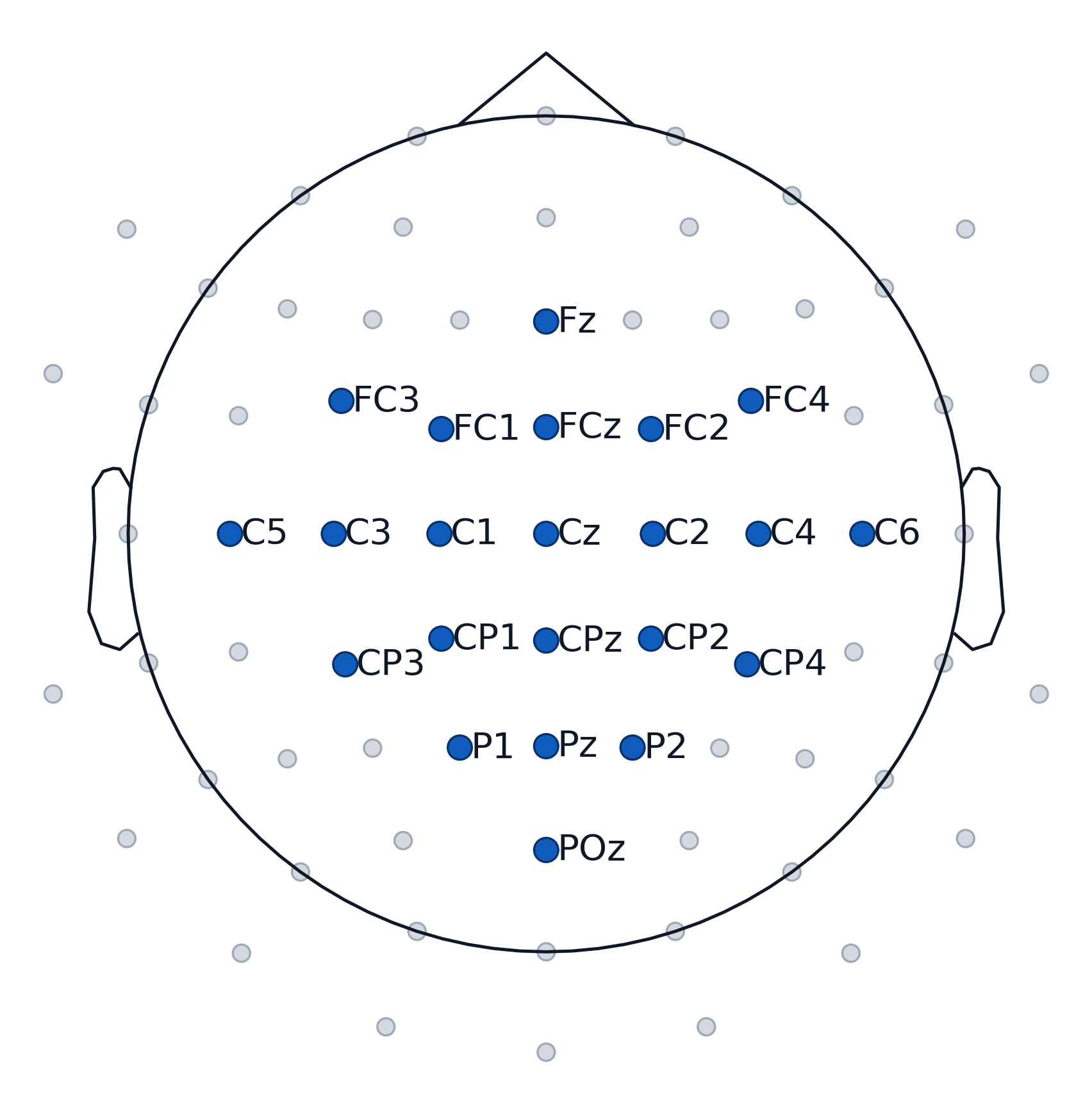}
        \caption{22-channel montage used for UET175 dataset (S22 channel subset)}
        \label{fig:easycap_uet22}
    \end{subfigure}

    \vspace{0.8em}

    \begin{subfigure}[t]{0.485\linewidth}
        \centering
        \includegraphics[width=\linewidth]{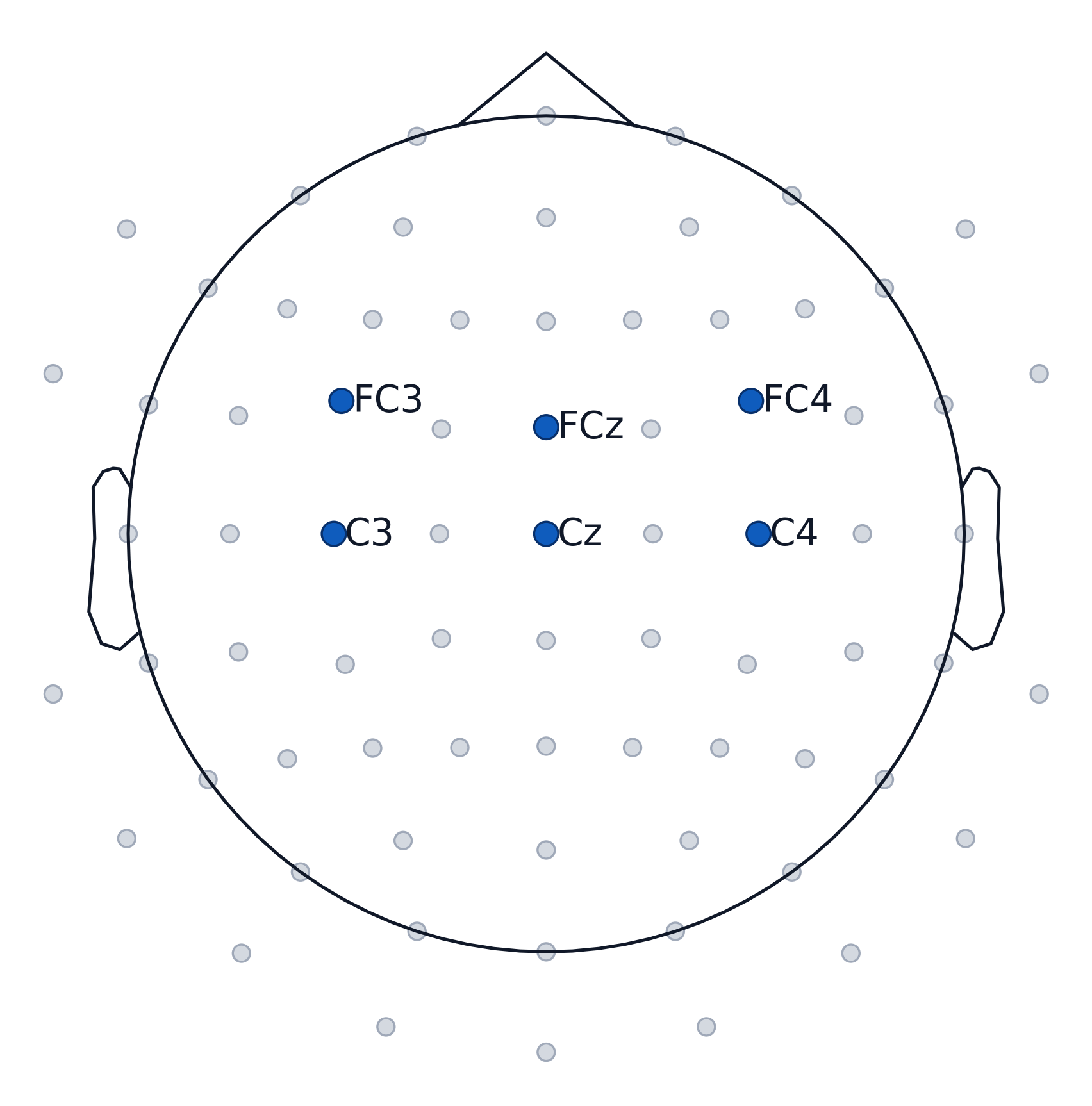}
        \caption{FC+C channel subset}
        \label{fig:easycap_fcc}
    \end{subfigure}
    \hfill
    \begin{subfigure}[t]{0.485\linewidth}
        \centering
        \includegraphics[width=\linewidth]{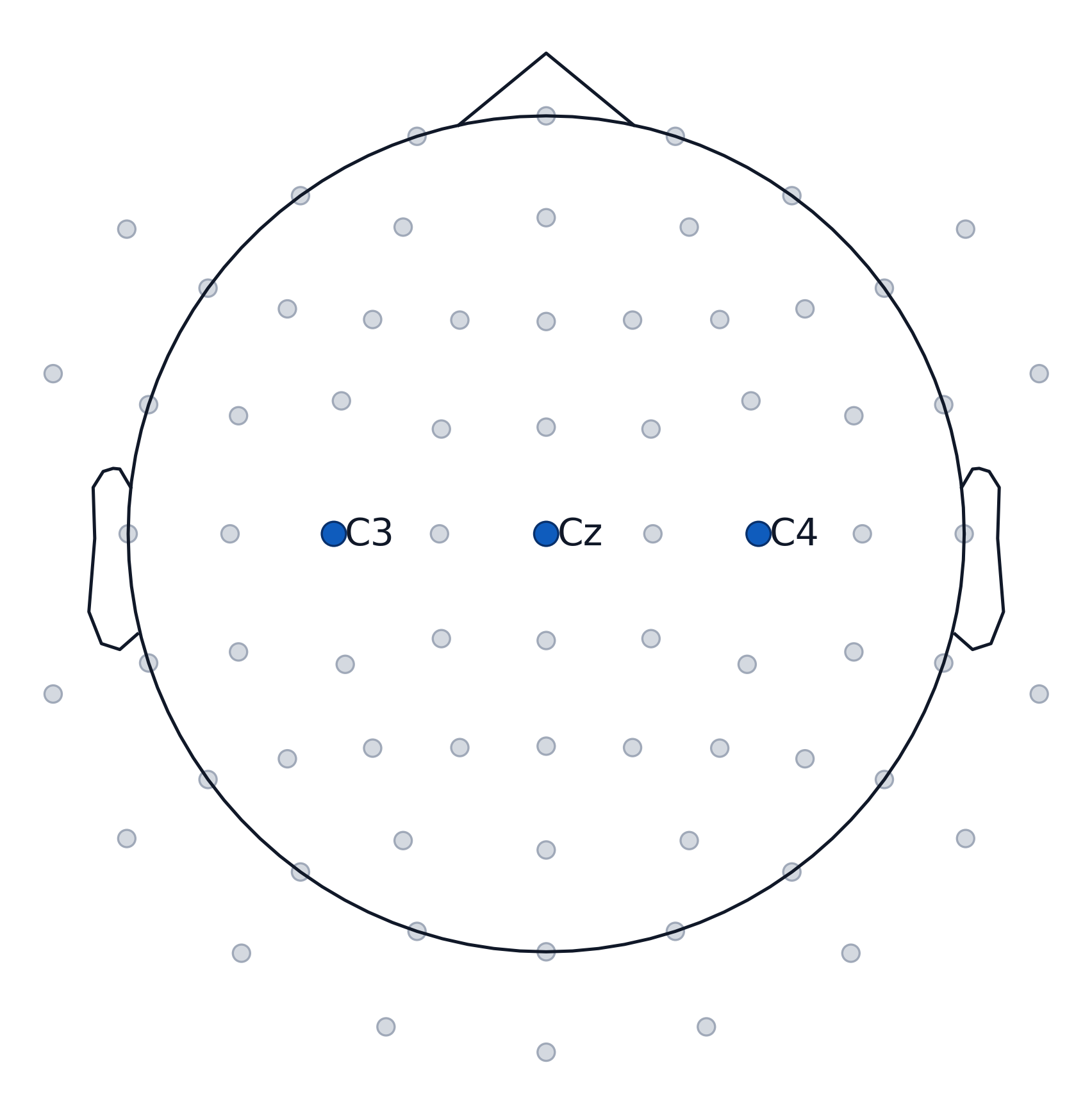}
        \caption{C-only channel subset}
        \label{fig:easycap_conly}
    \end{subfigure}

    \caption{EEG montages and channel subsets evaluated in this study. The 10--10 International EEG system template is shown with all channel positions in the background and the active subset highlighted for each layout.}
    \label{fig:channel_ablation_layouts}
\end{figure}

To test whether the spatial input configuration affects stroke MI decoding, we perform channel ablations using REVE-base LoRA under the same stroke LOOCV protocol. Three channel subsets are evaluated:

\begin{itemize}
    \item \textbf{S22}: the 22-channel montage used for the UET175 dataset (Fig. \ref{fig:easycap_uet22}).
    \item \textbf{FC+C}: including FC3, FCZ, FC4, C3, CZ, and C4 (Fig. \ref{fig:easycap_fcc}).
    \item \textbf{C-only}: including C3, CZ, and C4 (Fig. \ref{fig:easycap_conly}).
\end{itemize}

The goal is to determine whether restricting the model to motor-region channels improves decoding robustness, especially for low-performing subjects.

\subsubsection{Time-Window Ablation}

To examine whether MI information is concentrated in specific temporal segments, we perform a time-window ablation using REVE-base LoRA under the stroke LOOCV protocol. The reference setting uses the full 0--4\,s trial. We additionally evaluate 0--2.5\,s, 0.3--2.5\,s, 0.5--2.5\,s, and 1.0--3.0\,s windows.

To preserve the REVE input structure, each selected window is zero-padded back to the original $(C,800)$ shape before being passed into the model. Thus, all time-window conditions preserve the same patch structure while restricting the available temporal information.



\subsubsection{Clinical and ERD-Based Failure Analysis}

To interpret subject-level failure cases, we compare the two LOOCV low-performing subjects, ID01 and ID02, with lesion-matched high-performing controls, ID09 and ID05, respectively. Clinical matching is based on lesion location and available severity measures, including NIHSS, mRS, and muscle strength.

We further perform ERD analysis using channels C3, Cz, and C4 in the mu band (8--13\,Hz) and beta band (13--30\,Hz). ERD is a standard approach for characterizing task-related decreases in rhythmic sensorimotor EEG power during MI and movement-related paradigms \cite{pfurtscheller1999erd}. In this study, ERD is computed from raw CSV signals using a $-1$ to 0\,s baseline and a 0 to +4\,s task window:
\[
\mathrm{ERD}(\%) = \frac{P_{\mathrm{task}} - P_{\mathrm{baseline}}}{P_{\mathrm{baseline}}} \times 100.
\]
Negative values indicate desynchronization. For selected summary comparisons, we also report mean ERD values over the 0.5--3.5\,s interval.

In addition, we include a three-level ERD comparison among one healthy EEGMMIDB reference subject, one high-performing stroke subject, and one low-performing stroke subject. The healthy reference subject is selected based on strong contralateral beta lateralization among scanned EEGMMIDB candidates. This analysis is used to compare canonical healthy MI patterns with preserved and failed stroke MI patterns.

\subsubsection{Zero-Shot Healthy-to-Stroke Transfer}

To test whether a model adapted on healthy EEG can be directly deployed on stroke EEG, we evaluate zero-shot transfer from EEGMMIDB to UET175 subset. The best-performing model is trained on EEGMMIDB and evaluated directly on all 30 stroke subjects without stroke-domain fine-tuning. The 22-channel UET175 input is mapped into the 64-channel EEGMMIDB space, with missing channels zero-padded. The best-performing time window identified in the time-window ablation above was used for this zero-shot evaluation.

\section{Results}

\subsection{Dataset Summary After Preprocessing}

The final standard-preprocessed EEGMMIDB analytic dataset contains 109 subjects and 4,635 trials of unilateral fist MI, including 2,299 right-hand trials and 2,336 left-hand trials. Each trial is represented as a 4.0\,s EEG segment with shape $(64, 800)$ sampled at 200\,Hz.

For the ASR-preprocessed EEGMMIDB variant used in the head-only ablation, the dataset contains 4,101 trials, comprising 2,039 right-hand trials and 2,062 left-hand trials. Standard-preprocessed EEGMMIDB inputs were normalized by dividing microvolt values by 100, whereas ASR-preprocessed inputs were divided by 40 to account for the altered signal scale after artifact subspace reconstruction.

The final processed binary UET175 subset contains 1,281 accepted trials across 30 subjects, including 641 right-hand trials and 640 left-hand trials. Each trial is represented as a 4.0\,s EEG segment with shape $(22, 800)$ sampled at 200\,Hz. The retained trials were obtained after aligning the released event timestamp files with the original trial indices and correcting the left/right MI label assignment. All 30 subjects were retained; no subject was excluded based on decoding performance.

\subsection{Primary Analysis: Head-Only Baseline and LoRA Adaptation}
\subsubsection{Head-Only Baselines Show Limited Stroke Transferability}

We first evaluated LaBraM-base, REVE-base, and REVE-large in the head-only setting (Table \ref{tab:model_performance}). On EEGMMIDB, head-only performance followed the expected model-capacity trend, with LaBraM-base achieving 0.546$\pm$0.010 accuracy, REVE-base achieving 0.619$\pm$0.018, and REVE-large achieving 0.697$\pm$0.023. These results indicate that pretrained EEG representations provide useful features for healthy MI decoding even when the backbone is frozen.

However, performance dropped substantially on the UET175 subset. In the stroke head-only setting, LaBraM-base, REVE-base, and REVE-large all remained close to chance level, achieving 0.484$\pm$0.016, 0.495$\pm$0.022, and 0.492$\pm$0.014 accuracy, respectively. This indicates that frozen pretrained features alone are insufficient for robust stroke MI decoding and suggests substantial domain shift between healthy EEG and post-stroke EEG. With LoRA adaptation, the benefit was strongly architecture-dependent: LaBraM-base remained near chance at 0.499$\pm$0.009 accuracy, whereas REVE-base improved to 0.847$\pm$0.194. This motivated the use of REVE-base LoRA as the main configuration for subsequent stroke analyses. This primary comparison used the 5-fold subject-wise evaluation protocol for model selection, whereas the subsequent LOOCV analysis was used to characterize subject-level robustness of the selected configuration.

\begin{table}[t]
\centering
\footnotesize
\caption{Comparison of classification accuracy across foundation model settings}
\label{tab:model_performance}
\begin{tabular}{llll}
\hline
Mode & Model & EEGMMIDB & UET175 Subset \\
\hline
Head-only & LaBraM & 0.546 $\pm$ 0.010 & 0.484 $\pm$ 0.016 \\
Head-only & REVE-base & 0.619 $\pm$ 0.018 & 0.495 $\pm$ 0.022 \\
Head-only & REVE-large & 0.697 $\pm$ 0.023 & 0.492 $\pm$ 0.014 \\
LoRA & LaBraM & \textbf{0.822 $\pm$ 0.017} & 0.499 $\pm$ 0.009 \\
LoRA & REVE-base & \textbf{0.957 $\pm$ 0.056} & \textbf{0.847 $\pm$ 0.194} \\
LoRA & REVE-large & \textbf{0.948 $\pm$ 0.050} & \textbf{0.806 $\pm$ 0.178} \\
\hline
\end{tabular}
\end{table}

\subsubsection{LoRA Adaptation Improves Healthy EEG Decoding but Shows Divergent Transferability on Stroke EEG}

We next applied LoRA fine-tuning to LaBraM-base and REVE-base (Table \ref{tab:model_performance}). On EEGMMIDB, LoRA substantially improved both models. LaBraM-base increased from 0.546 to 0.822 accuracy, while REVE-base increased from 0.619 to 0.957. This shows that lightweight adaptation can strongly improve downstream MI decoding when the target domain is healthy EEG.

On the UET175 subset, however, LoRA adaptation showed sharply different behavior across models. LaBraM-base increased only from 0.484 to 0.499 accuracy, remaining near chance. In contrast, REVE-base improved from 0.495 to 0.847 accuracy after LoRA adaptation. This result indicates that LoRA is not sufficient by itself; the pretrained representation must also remain adaptable under stroke-related domain shift.

As an additional model-capacity control, we also evaluated REVE-large with LoRA. REVE-large LoRA achieved 0.948 ± 0.050 accuracy on EEGMMIDB and 0.806 ± 0.178 accuracy on the UET175 subset. Although these results were strong, they did not exceed REVE-base LoRA, which achieved 0.957 ± 0.056 on EEGMMIDB and 0.847 ± 0.194 on stroke EEG. This suggests that increasing model capacity alone does not necessarily improve downstream adaptation under stroke-domain shift. Instead, REVE-base provided the best performance-cost tradeoff among the evaluated configurations.

Together, these results show that successful adaptation on healthy EEG does not guarantee successful adaptation on stroke EEG. REVE-base provides substantially stronger transferability than LaBraM-base under the same lightweight adaptation strategy.
Based on the primary analysis, REVE-base with LoRA was selected as the main configuration for the subsequent analyses because it provided the strongest practical combination of stroke-decoding performance and lightweight adaptability.

\subsection{Secondary Analyses Using the Best Primary Configuration}
The following secondary analyses focus on the best configuration identified in the primary analysis, namely REVE-base with LoRA. Unless otherwise noted, the purpose of these analyses is not further model selection, but to examine the robustness, transfer behavior, and sensitivity of this strongest practical stroke-decoding setting.

\subsubsection{ASR-Based Preprocessing Degrades Performance in All Evaluated Settings}

We evaluated ASR-based preprocessing as a secondary ablation to determine whether stronger artifact removal improves downstream decoding in the evaluated follow-up settings. As shown in Table \ref{tab:asr_impact}, ASR reduced performance relative to the standard preprocessing pipeline across all tested configurations.

On EEGMMIDB, ASR decreased performance in all evaluated settings. For head-only baselines, LaBraM-base decreased from 0.546$\pm$0.010 to 0.503$\pm$0.016, REVE-base decreased from 0.619$\pm$0.018 to 0.540$\pm$0.012, and REVE-large decreased from 0.697$\pm$0.023 to 0.607$\pm$0.021. In addition, for REVE-base LoRA, ASR reduced accuracy from 0.957$\pm$0.056 to 0.896$\pm$0.060. These decreases were also reflected in balanced accuracy, Cohen's $\kappa$, and weighted F1-score.

On the UET175 subset, ASR also degraded LoRA-adapted performance. LaBraM-base decreased from 0.499$\pm$0.009 to 0.463$\pm$0.024, while REVE-base decreased from 0.847$\pm$0.194 to 0.719$\pm$0.243. Thus, in the present pipeline, ASR did not improve any evaluated condition. This finding should not be interpreted as evidence that ASR is universally harmful; rather, it indicates that ASR altered the signal distribution in a way that was unfavorable for these pretrained foundation-model pipelines.

\begin{table}[t]
\centering
\footnotesize
\caption{Mean classification accuracy for the ASR-based preprocessing ablation across foundation model settings}
\label{tab:asr_impact}
\begin{tabular}{m{0.1\linewidth}m{0.12\linewidth}m{0.18\linewidth}m{0.1\linewidth}m{0.1\linewidth}m{0.1\linewidth}}
\hline
Mode & Model & Dataset & non-ASR & ASR & Delta (pp) \\
\hline
Head-only & LaBraM-base & EEGMMIDB & 0.546 & 0.503 & -4.3 \\
Head-only & REVE-base & EEGMMIDB & 0.619 & 0.540 & -7.9 \\
Head-only & REVE-large & EEGMMIDB & 0.697 & 0.607 & -9.0 \\
\hline
LoRA & REVE-base & EEGMMIDB & 0.957 & 0.896 & -6.1 \\
LoRA & REVE-base & UET175 subset & 0.847 & 0.719 & -12.8 \\
\hline
\end{tabular}
\end{table}

\subsubsection{LOOCV Reveals a Small Low-Performing Subject Tail}

\begin{figure}[h]
    \centering
    \includegraphics[width=0.95\columnwidth]{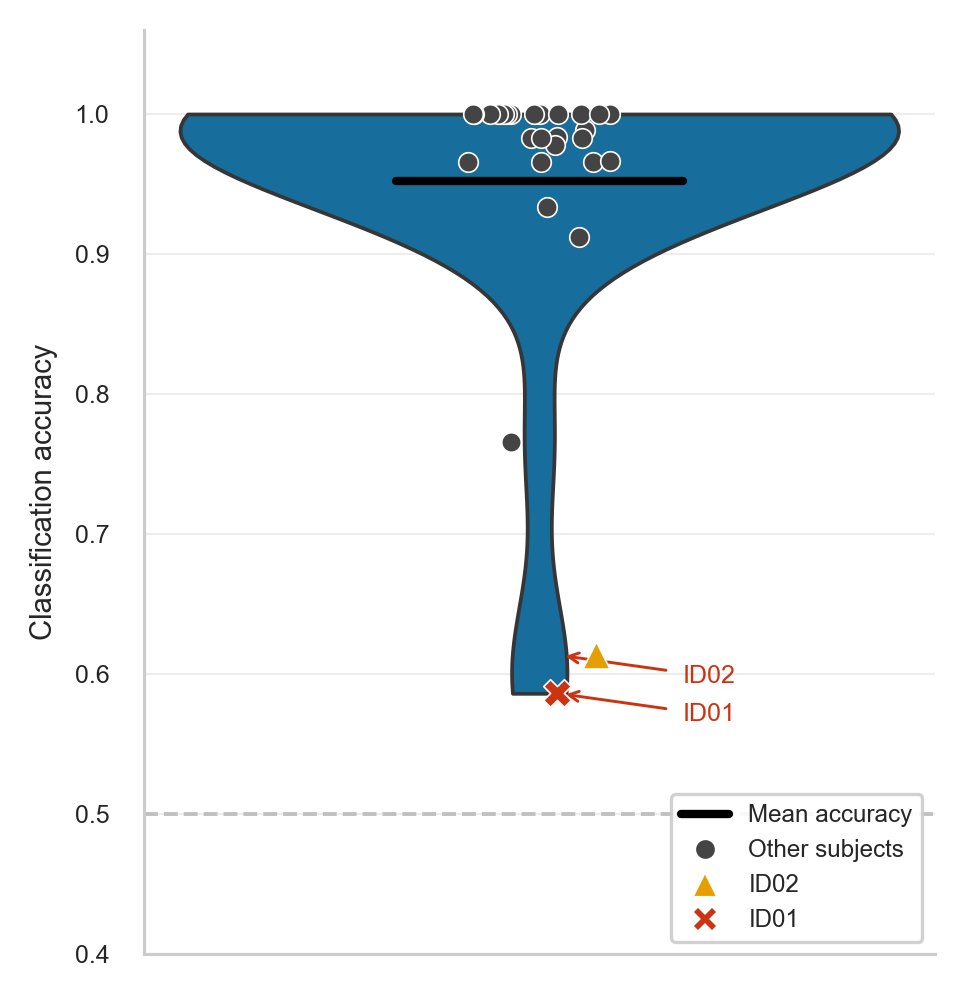}
    \caption{
    Subject-level LOOCV accuracy of REVE-base LoRA for for the 30 participants in the UET175 subset.
    The distribution shows a high-performing majority with a small low-performing tail.
    ID01 and ID02 were the two lowest-performing subjects, while most subjects achieved high decoding accuracy.
    The dashed line indicates chance level.
    }
    \label{fig:loocv_violin}
\end{figure}

We next evaluated the best-performing stroke configuration, REVE-base LoRA, using subject-level LOOCV (Fig. \ref{fig:loocv_violin}). Across 30 stroke subjects, the model achieved a mean accuracy of 0.9524 with a standard deviation of 0.1042. The median accuracy was 0.9945, and subject-wise accuracy ranged from 0.586 to 1.000.

The subject-wise distribution remained strongly skewed toward high performance, but a small low-performing tail was still present. ID01 and ID02 were the two lowest-performing subjects, with accuracies of 0.586 and 0.614, respectively. In contrast, 27 of 30 subjects achieved accuracy above 0.90, and 15 subjects achieved perfect accuracy. Thus, the overall variance was not distributed uniformly across the cohort, but was influenced mainly by a small number of lower-performing subjects.

This result shows that average performance alone is insufficient for evaluating clinical robustness. Although REVE-base LoRA performs strongly at the cohort level, subject-level analysis reveals a small subset of patients for whom the model remains unreliable.

\subsubsection{Channel Ablation Reveals Subject-Specific Spatial Sensitivity}

\begin{figure}[h]
    \centering
    \includegraphics[width=0.95\columnwidth]{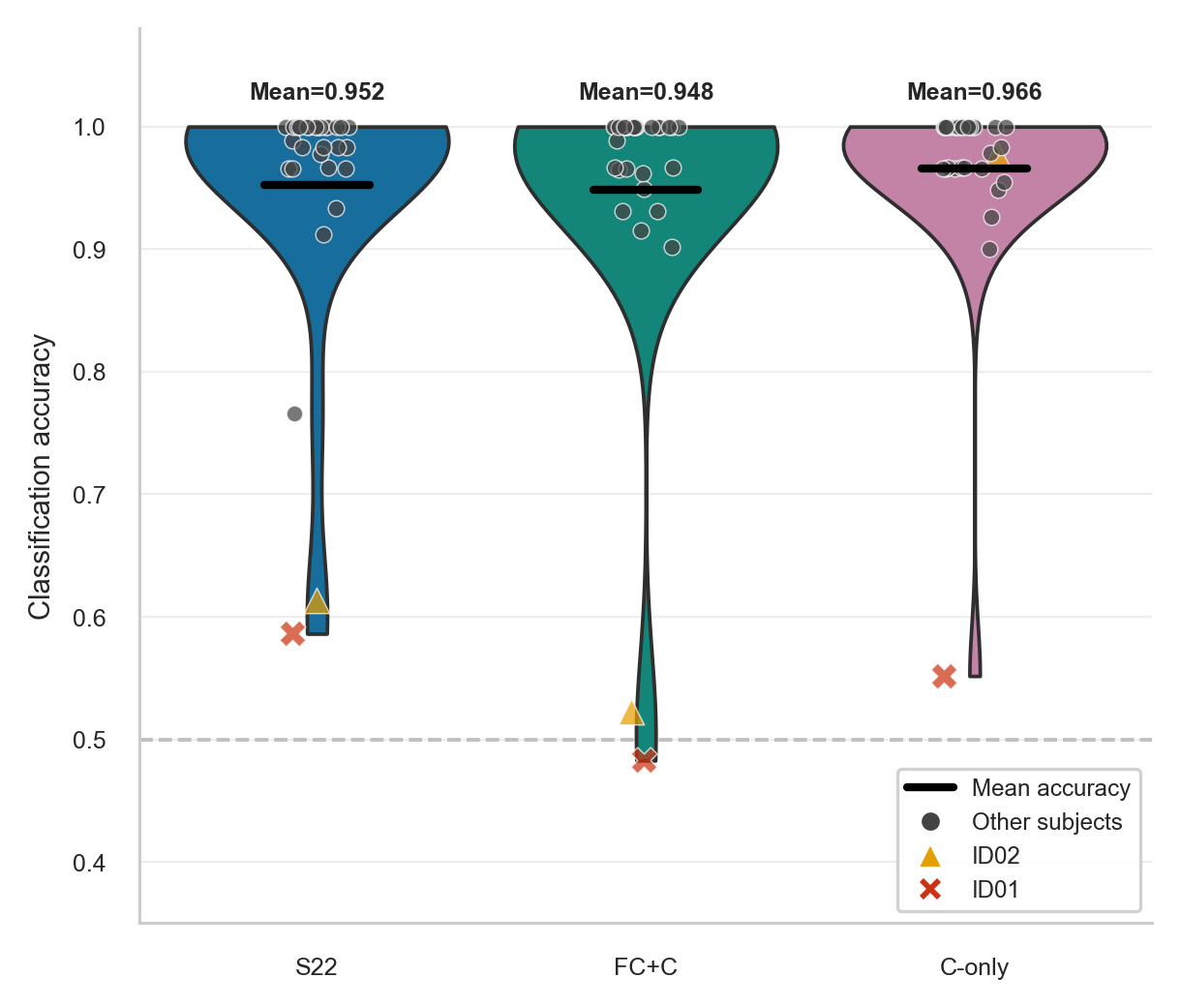}
    \caption{
    Channel ablation results for REVE-base LoRA under stroke LOOCV.
    The C-only channel subset achieved the highest numerical mean accuracy, but the improvement over the S22 layout was not statistically significant.
    ID02 improved under C-only input, whereas ID01 did not benefit from channel restriction.
    The dashed line indicates chance level.
    }
    \label{fig:channel_ablation}
\end{figure}

To assess whether decoding performance depended on spatial input configuration, we compared the S22 subset with two reduced motor-region subsets: C-only (C3, CZ, C4) and FC+C (FC3, FCZ, FC4, C3, CZ, C4) (Fig. \ref{fig:channel_ablation}). The S22 setting achieved a mean accuracy of 0.9524 with a standard deviation of 0.1042 and Cohen's $\kappa=0.9041$. The C-only channel subset achieved a numerically higher mean accuracy of 0.9661 with a lower standard deviation of 0.0808 and Cohen's $\kappa=0.9322$, but this difference relative to the S22 setting was not statistically significant in a paired subject-wise comparison (paired t-test, $p=0.37$). The FC+C channel subset achieved a mean accuracy of 0.9483, standard deviation of 0.1225, and Cohen's $\kappa=0.8958$, which was comparable to but slightly below the full-channel baseline.

The apparent improvement under C-only was driven primarily by a small number of subjects. In particular, ID02 improved from 0.6136 to 0.9773, and ID05 improved from 0.766 to 0.979 under C-only input. However, only 7 subjects improved with C-only, whereas 10 subjects showed modest decreases and 13 remained unchanged. ID01 did not benefit from C-only and decreased from 0.5862 to 0.5517. In contrast, the FC+C channel subset did not improve the two lowest-performing subjects: ID01 decreased to 0.4828 and ID02 decreased to 0.5227.

These results suggest that central motor channels may be beneficial for selected subjects with decoding difficulty, but C-only should not be interpreted as a universal replacement for the S22 layout. Adding frontal-central channels did not provide additional benefit in this setting. Overall, the channel-ablation results support subject-specific spatial sensitivity, while the S22 montage remains the more general default configuration for the stroke cohort.

\subsubsection{Time-Window Ablation Yields the Highest Mean Accuracy for the \texorpdfstring{0.5--2.5\,s}{0.5--2.5 s} Stroke MI Window}

We performed a time-window ablation to determine whether the most discriminative MI information was concentrated within a specific post-cue interval. The full 0--4\,s trial served as the baseline, achieving 0.9504$\pm$0.1241 accuracy and Cohen's $\kappa=0.9011$.

\begin{figure}[h]
    \centering
    \includegraphics[width=0.95\columnwidth]{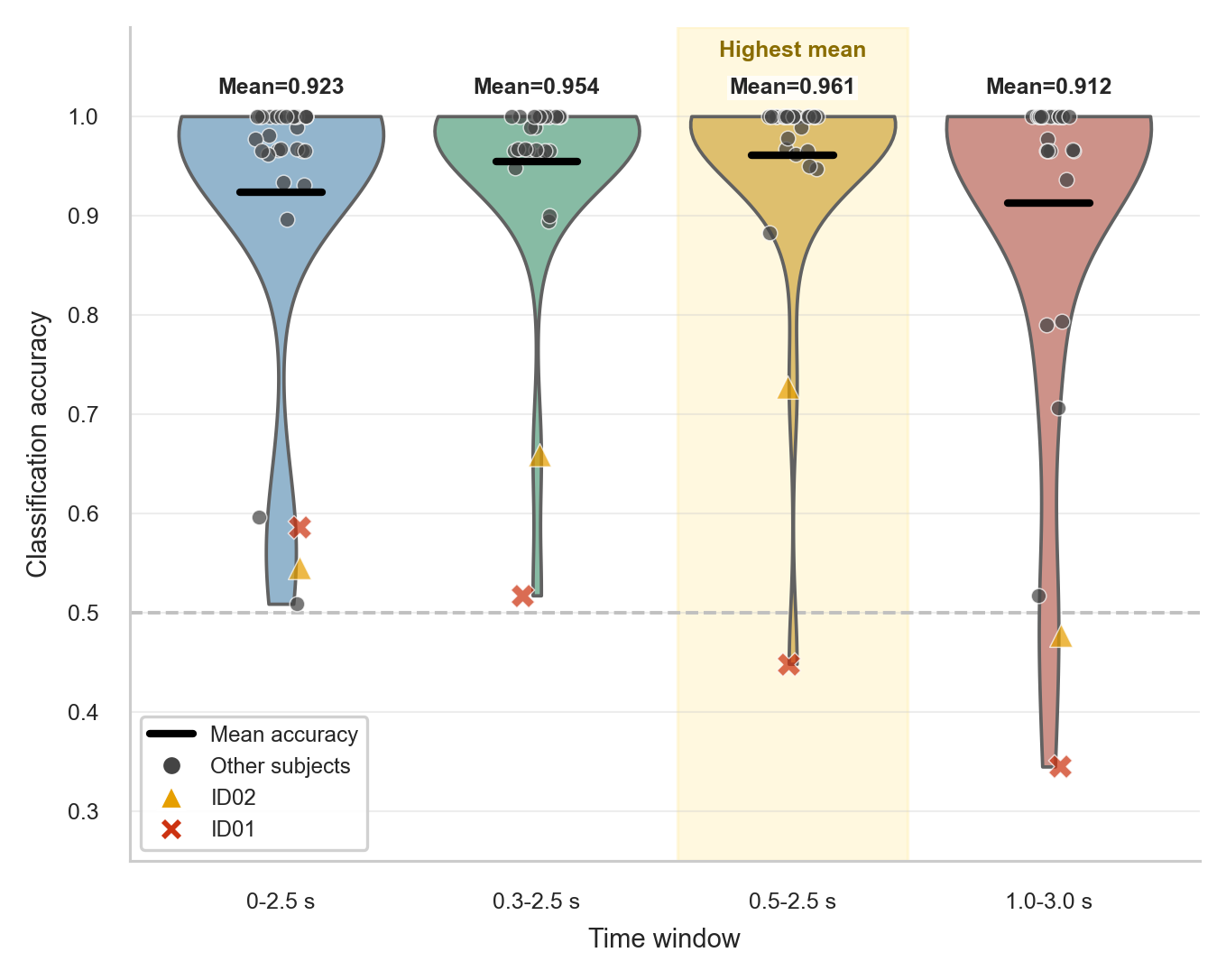}
    \caption{
    Time-window ablation results for REVE-base LoRA under stroke LOOCV.
    The 0.5--2.5\,s window achieved the highest mean accuracy among the tested temporal windows.
    However, subject-level responses were heterogeneous: ID02 improved under the 0.5--2.5\,s window, whereas ID01 did not benefit from temporal cropping.
    The dashed line indicates chance level.
    }
    \label{fig:timewin_ablation}
\end{figure}

In Fig. \ref{fig:timewin_ablation}, among all tested windows, 0.5--2.5\,s achieved the highest mean accuracy and Cohen's $\kappa$, with mean accuracy of 0.9605, standard deviation of 0.1089, and Cohen's $\kappa=0.9205$. This window achieved the highest mean accuracy and Cohen's $\kappa$, suggesting that the most discriminative stroke MI information is concentrated in an early post-cue interval rather than uniformly distributed across the full trial.

The time-window effect was especially important for ID02. Under the full 4\,s window, ID02 achieved 0.6136 accuracy. Under the 0.5--2.5\,s window, ID02 improved to 0.7273. In contrast, ID01 did not benefit from temporal cropping: its accuracy decreased from 0.5862 under the full 4\,s window to 0.4483 under the 0.5--2.5\,s window. Thus, temporal refinement can improve some low-performing subjects, but its benefit is not universal.

The 0--2.5\,s window achieved 0.9234 mean accuracy and Cohen's $\kappa=0.8458$, while the 0.3--2.5\,s window achieved 0.9540 and Cohen's $\kappa=0.9078$. The 1.0--3.0\,s window performed worst among the tested cropped windows, with 0.9123 mean accuracy and Cohen's $\kappa=0.8245$. These results suggest that the 0--0.5\,s onset period may contain subject-specific information, but including it does not provide the best cohort-level performance.

\subsubsection{Clinical Matching Shows That Failure Is Not Explained Solely by Gross Lesion Severity}

We compared the two lowest-performing subjects with lesion-matched comparison subjects. ID01 was matched with ID09 because both had right temporal lobe infarcts and identical available severity measures, including NIHSS=2, mRS=2, and muscle strength 4/5. Despite this clinical similarity, ID01 remained the lowest-performing subject in the corrected LOOCV analysis, with 0.586 accuracy, whereas ID09 achieved 0.983 accuracy.

ID02 was matched with ID05 based on shared involvement of the left lentiform region and comparable clinical severity measures. ID02 presented with a left lentiform nucleus infarct, whereas ID05 presented with a left lentiform nucleus and internal capsule hemorrhage. ID02 achieved 0.614 accuracy, whereas ID05 achieved 0.766. 
These comparisons indicate that gross clinical severity and lesion location alone do not fully explain decoding difficulty.

\subsubsection{ERD Analysis Shows That Failure Reflects Poor Discriminability Rather Than Absence of MI Activity}

\begin{figure*}[!t]
    \centering
    \includegraphics[width=0.98\textwidth]{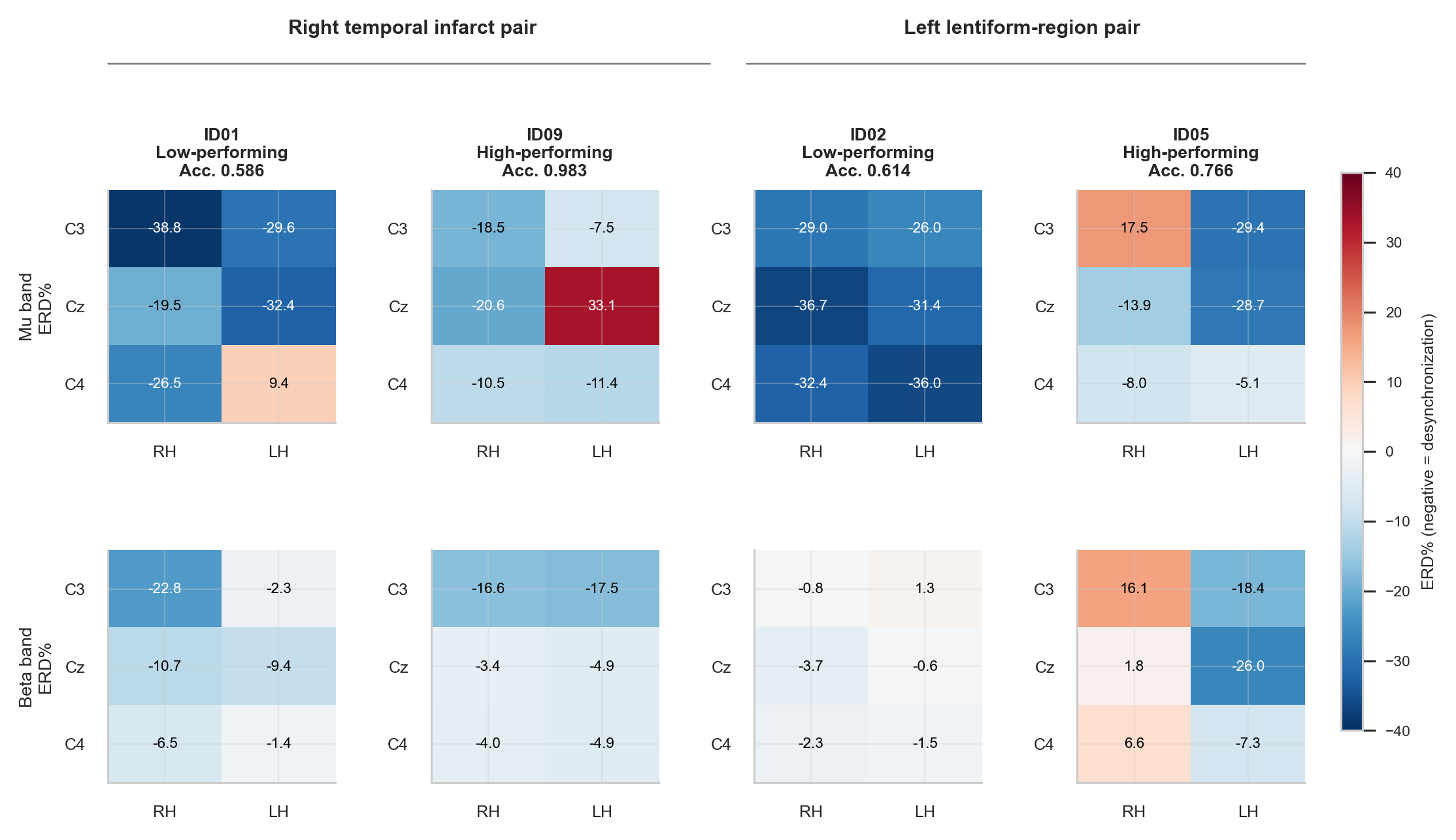}
    \caption{
    Matched-pair ERD comparison between low-performing stroke participants and clinically similar higher-performing comparison participants. The right temporal infarct pair comprises ID01 and ID09, with LOOCV accuracies of 0.586 and 0.983, respectively. The left lentiform-region pair comprises ID02 and ID05, with LOOCV accuracies of 0.614 and 0.766, respectively. ERD values were computed for C3, Cz, and C4 in the mu and beta bands during right-hand (RH) and left-hand (LH) motor imagery. Negative values, shown in blue, indicate event-related desynchronization, whereas positive values, shown in red, indicate event-related synchronization. Both low-performing participants retained mu-band ERD, but ID01 showed abnormal task-dependent spatial organization and ID02 showed nearly symmetric mu ERD with minimal beta modulation. The higher-performing comparison participants exhibited more distinct task-dependent spatial and frequency-band patterns.
    }
    \label{fig:erd_matched_pairs}
\end{figure*}

We analyzed ERD patterns in ID01 and ID02 together with their clinically matched higher-performing comparison participants, ID09 and ID05, respectively (Fig. \ref{fig:erd_matched_pairs}). Both low-performing participants exhibited substantial mu-band ERD, indicating that MI-related neural activity was not completely absent. However, their spatial distribution, task specificity, and frequency-band organization differed from those of the higher-performing comparison participants.

ID01 showed strong mu ERD during both MI tasks, but its spatial organization was abnormal: during left-hand imagery, C3 and Cz remained strongly desynchronized while C4 changed to positive ERS, and beta modulation was weak. ID09 showed a more distinct task-dependent mu configuration and retained beta desynchronization. ID02 exhibited a different failure pattern, with nearly symmetric mu ERD across right- and left-hand imagery and beta modulation close to zero. In contrast, ID05 showed pronounced task-dependent changes in both bands, including a transition from predominantly positive beta values during right-hand imagery to negative values during left-hand imagery.

These comparisons indicate that low decoding performance was not caused by a complete absence of MI-related activity. Instead, the two low-performing participants showed different forms of physiological non-discriminability: ID01 retained strong but abnormally organized modulation, whereas ID02 showed similar responses across the two tasks. Thus, successful decoding appeared to depend less on ERD magnitude alone than on stable, task-dependent spatial and frequency-band structure.

\subsubsection{Three-Level ERD Comparison Across Healthy, High-Performing Stroke, and Low-Performing Stroke Participants}

\begin{figure*}[h]
    \centering
    \includegraphics[width=0.80\textwidth]{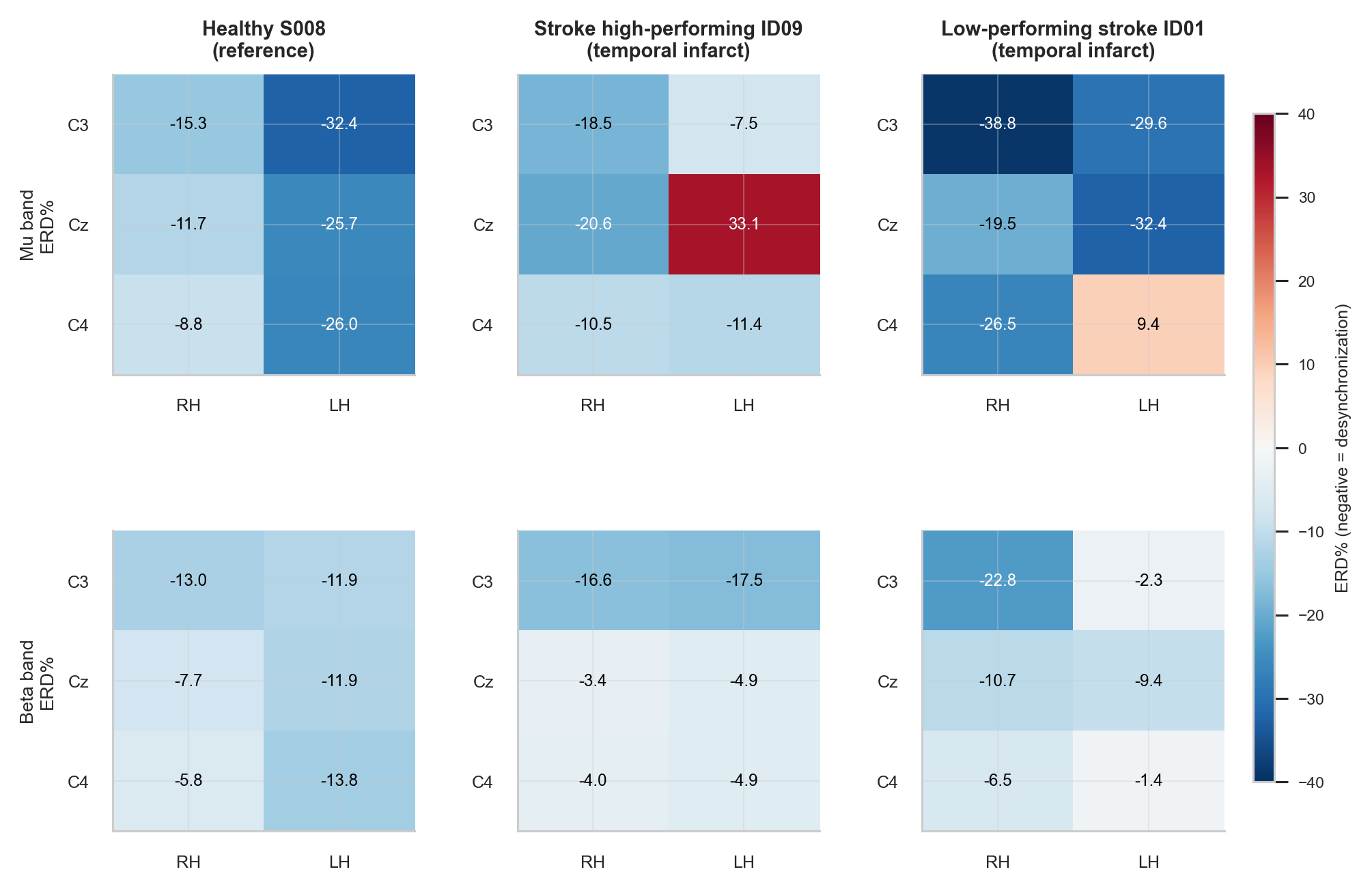}
    \caption{
    Three-level ERD comparison across the healthy reference participant S008, the high-performing stroke participant ID09, and the low-performing stroke participant ID01. ERD values were computed for C3, Cz, and C4 in the mu and beta bands during right-hand (RH) and left-hand (LH) motor imagery. Negative values are shown in blue and positive values in red. The comparison illustrates that substantial ERD magnitude may remain present in a low-performing participant despite weaker task-dependent spatial and frequency-band organization.
    }
    \label{fig:erd_full}
\end{figure*}

To provide an illustrative healthy reference, we compared EEGMMIDB participant S008 with the high-performing stroke participant ID09 and the low-performing stroke participant ID01 (Fig. \ref{fig:erd_full}). S008 was selected from the scanned healthy candidates based on comparatively clear task-related sensorimotor modulation and is not intended as a canonical template for the healthy population.

S008 showed relatively consistent desynchronization across the evaluated channels and frequency bands, whereas ID09 retained a distinguishable but more heterogeneous task-dependent spatial pattern. In contrast, ID01 exhibited substantial mu ERD but abnormal spatial organization and weak beta modulation during left-hand imagery. This comparison further indicates that ERD magnitude alone is insufficient for successful decoding; reliable classification depends on whether MI produces stable, task-dependent spatial and frequency-band structure.


\subsubsection{Zero-Shot Healthy-to-Stroke Transfer Fails Under Domain Shift}

To directly test whether a healthy-trained model can be deployed on stroke EEG without target-domain adaptation, we evaluated zero-shot transfer from EEGMMIDB to UET175 subset using REVE-base LoRA. The model was trained on EEGMMIDB and evaluated on all 30 stroke subjects without stroke fine-tuning.
For this zero-shot evaluation, we used the best-performing 0.5--2.5\,s window for MI task epochs.

Zero-shot transfer failed to generalize to the UET175 subset. Mean accuracy was 0.4638$\pm$0.0722, balanced accuracy was 0.4648$\pm$0.0715, Cohen's $\kappa$ was $-0.0700\pm0.1426$, and weighted F1-score was 0.4437$\pm$0.0732. These values are below or near chance, demonstrating that a model adapted on healthy EEG cannot be directly transferred to post-stroke EEG in this setting.

This result demonstrates poor transfer under the combined cross-dataset domain shift, which includes differences in population, task, montage, and acquisition hardware. It also highlights the necessity of target-domain adaptation: direct stroke LOOCV fine-tuning with REVE-base LoRA reaches substantially higher performance than zero-shot transfer.

\section{Discussion}

This study evaluated whether pretrained EEG foundation models can be adapted for binary stroke MI decoding under substantial healthy-to-stroke domain shift. 
The key finding is that strong performance on healthy EEG may not translate automatically to stroke EEG: although LoRA improved both LaBraM-base and REVE-base on EEGMMIDB, only REVE-base remained strongly adaptable on the UET175 stroke subset. 
Direct healthy-to-stroke transfer failed, showing that target-domain adaptation is necessary rather than optional. 
Secondary analyses revealed that ASR-based preprocessing degraded performance in all evaluated settings, suggesting that stronger artifact removal does not necessarily improve MI decoding performance of EEG foundation models. 
In addition, subject-level success with REVE-base LoRA did not eliminate subject-level failure cases, which remained visible under LOOCV and responded heterogeneously to spatial and temporal input restriction.

\subsection{Foundation Model Adaptation Under Stroke Domain Shift}
The primary analysis separates two questions that are often conflated: whether a pretrained model encodes useful MI features at all, and whether those features remain adaptable under clinical domain shift.
The strong improvement from LoRA on EEGMMIDB indicates that pretrained EEG foundation models contain useful representations for MI decoding that can be effectively unlocked through lightweight adaptation. 
On the UET175 subset, however, head-only performance for all three evaluated models remained near chance, and LaBraM-base remained near chance even after LoRA.
In contrast, REVE-base improved substantially under the same lightweight adaptation strategy. 
This pattern suggests that pretraining alone is insufficient; what matters in this setting is whether the pretrained representation remains compatible with stroke EEG after differences in population, montage, hardware, and acquisition conditions are introduced.

The REVE-large capacity-control result sharpens this interpretation. A larger backbone did not outperform REVE-base under LoRA, so the practical advantage here is not simply model scale. This may reflect the limited size and high heterogeneity of the UET175 subset, where a larger pretrained backbone is not necessarily easier to adapt through low-rank updates. Among the evaluated configurations, REVE-base provided the best performance-efficiency tradeoff for stroke MI decoding.

The zero-shot transfer experiment further supports this interpretation. A REVE-base LoRA model trained on EEGMMIDB failed when evaluated directly on UET175 stroke subjects, with performance below chance and negative Cohen's $\kappa$. 


Together, the LoRA and zero-shot results distinguish pretrained representation quality from clinical transferability, an important distinction given the limited and heterogeneous evaluations currently used for EEG foundation models \cite{kuruppuEEGFoundationModels2026}. In our experiments, strong performance on healthy EEG did not ensure zero-shot generalization to post-stroke EEG. This finding is consistent with evidence that stroke-related lesions alter MI-associated EEG patterns and that adapting healthy-pretrained models to stroke data can substantially improve decoding performance \cite{park2016eeg_lesion,nagarajan2024healthy_to_stroke}. Healthy-trained MI decoders therefore should not be assumed ready for stroke rehabilitation without target-domain validation and, where necessary, calibration or fine-tuning. Evaluation should also examine subject-level robustness, because clinical studies have found marked inter-patient variability and chance-level performance in subsets of stroke participants \cite{ang2011large_clinical,shu2018bci_inefficient}.

\subsection{Preprocessing Sensitivity of EEG Foundation Models}

The ASR-based preprocessing reduced performance across all evaluated follow-up configurations, including head-only models on EEGMMIDB and LoRA-adapted models on the UET175 stroke subset. Within the scope of this study, the most plausible interpretation is not that ASR is universally harmful, but that it changed the signal distribution in a way that was unfavorable for these pretrained pipelines. Foundation models are sensitive not only to artifact level, but also to the statistical structure of the input they receive. In this context, stronger cleaning did not create a more useful representation space for downstream adaptation.

This finding is consistent with broader evidence that automated EEG cleaning does not uniformly improve the information available for downstream analysis. Delorme reported that ASR-based rejection of contaminated segments did not reliably improve an ERP condition-separability measure once the accompanying loss of trials and statistical power was considered \cite{delorme2023eeg}. Although that study evaluated ERP analyses rather than MI decoding or pretrained models, it similarly demonstrates that preprocessing effectiveness depends on the analysis objective and evaluation metric.

This finding is important because preprocessing choices are often assumed to improve EEG decoding by removing noise. For pretrained EEG foundation models, preprocessing may introduce an additional source of sensitivity because it can alter the statistical and spatiotemporal structure of the model input. Aggressive cleaning may remove or attenuate components that are useful for MI decoding, or may shift the input distribution away from what the pretrained model expects. 
Therefore, for EEG foundation models in particular, preprocessing choices should be validated empirically on the target clinical pipeline instead of being inherited uncritically from conventional EEG workflows.

\subsection{Subject-Level Robustness and Heterogeneous Failure Modes}

The LOOCV analysis showed strong overall performance with REVE-base LoRA, but also revealed that the average accuracy was not representative of all subjects. Most held-out stroke subjects were decoded with high accuracy, yet a small low-performing tail remained. 
This matters because cohort means alone can obscure clinically important failure cases. At the same time, the LOOCV analysis here should be interpreted as within-cohort robustness analysis rather than external validation.
 
The follow-up channel and time-window ablations showed that the two lowest-performing subjects behaved differently. ID02 showed sensitivity to input selection, with a large improvement under the C-only EEG channel subset and a more modest improvement under the 0.5--2.5\,s time window. However, the C-only effect was not consistent across the cohort and did not reach statistical significance at the group level. This suggests that ID02 retained decodable MI information that may be more accessible under selected spatial or temporal input settings. In contrast, ID01 remained difficult across channel and time-window configurations and did not benefit from the best-performing spatial or temporal refinements.

These findings support the view that stroke MI decoding failure is heterogeneous. Some low-performing cases may be improved by better selecting the spatial or temporal input region, whereas others may reflect more fundamental disruption of task-discriminative neural organization. This distinction is important for rehabilitation BCI because it suggests that model failure should not be treated as a single phenomenon. Instead, future systems may need subject-specific diagnostic analyses to determine whether poor decoding is due to recoverable signal dilution or to more persistent physiological non-discriminability.

\subsection{Physiological Interpretation From ERD Analysis}

The ERD analyses provide additional context for the subject-level decoding results. Both ID01 and ID02 showed mu-band ERD, indicating that MI-related neural activity was not completely absent. However, their ERD patterns were not strongly task-discriminative. ID01 showed abnormal lateralization and weak task selectivity, while ID02 showed nearly symmetric mu desynchronization and minimal beta modulation. In contrast, matched high-performing controls showed clearer task-specific ERD patterns and stronger beta modulation. Altered or highly variable ERD lateralization has also been reported after stroke, reflecting lesion-dependent changes in the spatial organization of motor-related activity \cite{kaiser2012stroke_erd,braun2017mi_impairment}.

The three-level comparison with a healthy reference subject further clarifies this point. The healthy reference subject showed a more organized and interpretable task-related ERD pattern, the high-performing stroke subject retained task-specific organization, and the low-performing stroke subject showed large but non-specific ERD. This suggests that ERD magnitude alone might not be sufficient for decoding. Prior work likewise indicates that ERD strength does not necessarily track MI-BCI accuracy, whereas distinct task-dependent ERD/ERS patterns can improve single-trial discrimination \cite{kwon2018erd_performance,pfurtscheller2006mu_classification}. What matters for classification is whether the neural response contains stable, spatially organized, and task-specific information.

This interpretation helps explain why a subject may show strong neural activation but still perform poorly in classification. A decoder trained to distinguish left- versus right-hand MI requires consistent class-specific structure. If both classes produce similar or abnormally lateralized responses, then the model may fail even when overall ERD magnitude is large. Studies in patients with hemiplegia have similarly reported that comparable cortical-activation measures can coexist with differences in BCI accuracy, underscoring that activation strength and decodability are related but non-equivalent properties \cite{chen2021erd_accuracy}. Therefore, physiological interpretability is important for understanding when model failure reflects absence of engagement versus poor class separability.

\subsection{Implications for Rehabilitation BCI}

These results have several implications for rehabilitation BCI applications. First, pretrained EEG foundation models should not be evaluated only on healthy benchmark datasets such as EEGMMIDB if the intended application is stroke rehabilitation. Healthy-domain performance overestimates clinical transferability, as shown by the failed zero-shot transfer result. Second, lightweight target-domain adaptation is necessary, but model choice remains important: REVE-base showed substantially stronger adaptability than LaBraM-base under the same LoRA strategy. Third, subject-level robustness should be part of clinical BCI evaluation. Average accuracy can hide clinically important outliers, and those outliers may require different interventions such as temporal-window refinement, subject-specific channel selection, or alternative feedback paradigms.

In addition, the findings from time-window and channel ablation analyses suggest that spatial and temporal input selection should be interpreted as sensitivity analyses rather than as definitive universal preprocessing choices. For future real-time or pseudo-online rehabilitation systems, decoder calibration should include not only model fine-tuning, but also subject-specific checks of temporal informativeness, spatial sensitivity, and physiological discriminability.

\subsection{Limitations}

This study has several limitations. First, the imagined movements were not fully matched across datasets: EEGMMIDB involved unilateral fist opening-and-closing imagery, whereas UET175 involved lifting the left or right arm. Although both were reduced to binary left-versus-right classification, they differ in motor content and task paradigm inherently. Consequently, the observed healthy-to-stroke performance gap cannot be attributed exclusively to stroke-related neurophysiological domain shift, as task mismatch may also contribute substantially. 
Second, although LOOCV provides a subject-level estimate of generalization within the stroke cohort, the stroke cohort is limited to 30 subjects, and the LOOCV analysis does not replace external validation on additional post-stroke cohorts. 
Third, the ASR ablation was targeted rather than exhaustive, so the preprocessing conclusion should remain limited to the evaluated configurations. 
Finally, although the channel and time-window ablations provide useful insight, they do not exhaustively search all possible spatial or temporal configurations. 



\section{Conclusion}

This work investigated lightweight adaptation of EEG foundation models for stroke MI decoding under healthy-to-stroke domain shift. Across healthy EEG, LoRA substantially improved both LaBraM-base and REVE-base, indicating that pretrained EEG representations can be effectively adapted with a small number of trainable parameters. However, under stroke-domain shift, model transferability diverged sharply: LaBraM-base remained near chance, whereas REVE-base showed strong adaptation performance. The failed zero-shot transfer experiment further demonstrated that healthy-trained MI decoders cannot be directly deployed on stroke EEG without target-domain adaptation.


This work is limited to offline analyses, yet the present results still offer practical guidance for deployment-oriented research. 
In particular, these findings suggest that clinical deployment of foundation-model-based rehabilitation BCI requires more than strong performance on healthy benchmark datasets such as EEGMMIDB alone. Robust stroke MI decoding depends on target-domain adaptation, preprocessing-aware evaluation, subject-level robustness analysis, and physiological interpretation of low-performing cases. Among the evaluated models, REVE-base with LoRA provided the strongest practical adaptation strategy for stroke MI decoding. The channel and time-window analyses further suggest that spatial and temporal input choices can affect selected subjects, but should be interpreted as subject-specific sensitivity factors rather than universal optimal settings. Future work should validate these findings on additional stroke cohorts and evaluate whether the proposed adaptation strategy transfers to pseudo-online or real-time rehabilitation BCI settings.

\bibliographystyle{IEEEtran}
\bibliography{references}

@article{padfield2019mi_review,
  title={{EEG}-Based Brain-Computer Interfaces Using Motor-Imagery: Techniques and Challenges},
  author={Padfield, Natasha and Zabalza, Jaime and Zhao, Huimin and Masero, Vittorio and Ren, Jinchang},
  journal={Sensors},
  volume={19},
  number={6},
  pages={1423},
  year={2019}
}

@article{mulder2007mi_rehab,
  title={Motor imagery and action observation: cognitive tools for rehabilitation},
  author={Mulder, Theo},
  journal={Journal of Neural Transmission},
  volume={114},
  number={10},
  pages={1265--1278},
  year={2007}
}

@article{khan2020stroke_bci_review,
  title={Review on motor imagery based BCI systems for upper limb post-stroke neurorehabilitation: From designing to application},
  author={Khan, Muhammad Ahmed and Das, Rig and Iversen, Helle K and Puthusserypady, Sadasivan},
  journal={Computers in biology and medicine},
  volume={123},
  pages={103843},
  year={2020},
  publisher={Elsevier}
}

@article{liao2023mi_bci_rehab,
  title={Motor imagery brain–-computer interface rehabilitation system enhances upper limb performance and improves brain activity in stroke patients: A clinical study},
  author={Liao, Wenzhe and Li, Jiahao and Zhang, Xuesong and Li, Chen},
  journal={Frontiers in Human Neuroscience},
  volume={17},
  pages={1117670},
  year={2023}
}

@article{wang2024bci_ul_rehab,
  title={Rehabilitation with brain-computer interface and upper limb functional electrical stimulation in ischemic stroke: a randomized controlled trial},
  author={Wang, Anxin and Tian, Xue and Jiang, Di and Yang, Chengyuan and Xu, Qin and Zhang, Yifei and Zhao, Shaoqing and Zhang, Xiaoli and Jing, Jing and Wei, Ning and Wu, Yuqian and Lv, Wei and Yang, Banghua and Zang, Dawei and Wang, Yilong and Zhang, Yumei and Wang, Yongjun and Meng, Xia},
  journal={Brain and Spine},
  volume={4},
  pages={102837},
  year={2024}
}

@article{lawhern2018eegnet,
  title={{EEGNet: a compact convolutional neural network for {EEG}-based brain-computer interfaces}},
  author={Lawhern, Vernon J. and Solon, Amelia J. and Waytowich, Nicholas R. and Gordon, Stephen M. and Hung, Chou P. and Lance, Brent J.},
  journal={Journal of Neural Engineering},
  volume={15},
  number={5},
  pages={056013},
  year={2018}
}

@article{jiang2024labram,
  title={{Large Brain Model for Learning Generic Representations with Tremendous {EEG} Data in BCI}},
  author={Jiang, Wei-Bang and Zhao, Li-Ming and Lu, Bao-Liang},
  journal={arXiv preprint arXiv:2405.18765},
  year={2024}
}

@article{ouahidi2025reve,
  title={{REVE: A Foundation Model for {EEG}}},
  author={El Ouahidi, Yassine and Lys, Jonathan and Th{\"o}lke, Philipp and Farrugia, Nicolas and Pasdeloup, Bastien and Gripon, Vincent and Jerbi, Karim and Lioi, Giulia},
  journal={arXiv preprint arXiv:2510.21585},
  year={2025}
}

@article{hu2021lora,
  title={{LoRA: Low-Rank Adaptation of Large Language Models}},
  author={Hu, Edward J. and Shen, Yelong and Wallis, Phillip and Allen-Zhu, Zeyuan and Li, Yuanzhi and Wang, Shean and Wang, Lu and Chen, Weizhu},
  journal={arXiv preprint arXiv:2106.09685},
  year={2021}
}

@article{suzumura2024graphadapter,
  title={Graph Adapter of {EEG} Foundation Models for Parameter Efficient Fine Tuning},
  author={Suzumura, Toyotaro and Kanezashi, Hiroki and Akahori, Shotaro},
  journal={arXiv preprint arXiv:2411.16155},
  year={2024}
}

@article{lotey2024edora,
  title={{EEG}-Based Mental Imagery Task Adaptation via Ensemble of Weight-Decomposed Low-Rank Adapters},
  author={Lotey, Taveena and Verma, Aman and Roy, Partha Pratim},
  journal={arXiv preprint arXiv:2412.17818},
  year={2024}
}

@article{wang2025tcpl,
  title={{TCPL: task-conditioned prompt learning for few-shot cross-subject motor imagery {EEG} decoding}},
  author={Wang, Pengpai and Xie, Tiantian and Zhou, Yueying and Gong, Peiliang and Chan, Rosa H. M.},
  journal={Frontiers in Neuroscience},
  volume={19},
  pages={1689286},
  year={2025}
}

@article{carrara2024pseudo_online,
  title={{Pseudo-online framework for BCI evaluation: a MOABB perspective using various MI and SSVEP datasets}},
  author={Carrara, Igor and Papadopoulo, Theodore},
  journal={Journal of Neural Engineering},
  year={2024}
}

@article{acqualagna2016coadaptive,
  title={Large-Scale Assessment of a Fully Automatic Co-Adaptive Motor Imagery-Based Brain Computer Interface},
  author={Acqualagna, Laura and Botrel, Loic and Vidaurre, Carmen and K{\"u}bler, Andrea and Blankertz, Benjamin},
  journal={PLOS ONE},
  volume={11},
  number={2},
  pages={e0148886},
  year={2016}
}

@inproceedings{choi2024real_time_robot,
  title={On the Feasibility of {EEG}-based Motor Intention Detection for Real-Time Robot Assistive Control},
  author={Choi, Ho Jin and Das, Satyajeet and Peng, Shaoting and Bajcsy, Ruzena and Figueroa, Nadia},
  booktitle={2024 IEEE International Conference on Robotics and Automation (ICRA)},
  year={2024}
}

@article{eegmmidb_physionet,
  author  = {Schalk, Gerwin},
  title   = {{{EEG} Motor Movement/Imagery Dataset}},
  journal = {{PhysioNet}},
  year    = {2009},
  month   = sep,
  note    = {Version 1.0.0},
  doi     = {10.13026/C28G6P},
  url     = {https://doi.org/10.13026/C28G6P}
}

@article{uet175_dataset,
  author  = {Ma Thi, C. and Nguyen The, H.-A. and Nguyen Minh, K. and Vu Thanh, L. and Nguyen Dinh, H. and Huynh Thi, N.-Y. and Ha Thi, T.-H. and Hoang Tien, T.-N. and Au Dao, D.-T. and Nguyen Hoang, K.-L. and Huynh Kha, V. and Le Hoang, T.-L.},
  title   = {{{UET175}: {EEG} dataset of motor imagery tasks in Vietnamese stroke patients}},
  journal = {Frontiers in Neuroscience},
  volume  = {19},
  pages   = {1580931},
  year    = {2025},
  doi     = {10.3389/fnins.2025.1580931}
}

@article{mullen2015asr,
  author  = {Mullen, Tim R. and Kothe, Christian A. E. and Chi, Yu Mike and Ojeda, Alejandro and Kerth, Trevor and Makeig, Scott and Jung, Tzyy-Ping and Cauwenberghs, Gert},
  title   = {{Real-Time Neuroimaging and Cognitive Monitoring Using Wearable Dry {EEG}}},
  journal = {IEEE Transactions on Biomedical Engineering},
  volume  = {62},
  number  = {11},
  pages   = {2553--2567},
  year    = {2015},
  doi     = {10.1109/TBME.2015.2481482}
}

@article{pfurtscheller1999erd,
  author  = {Pfurtscheller, Gert and Lopes da Silva, Fernando H.},
  title   = {{Event-related {EEG}/MEG synchronization and desynchronization: basic principles}},
  journal = {Clinical Neurophysiology},
  volume  = {110},
  number  = {11},
  pages   = {1842--1857},
  year    = {1999},
  doi     = {10.1016/S1388-2457(99)00141-8}
}

@misc{banvilleNeuralBenchUnifyingFramework2026,
  title = {{{NeuralBench}}: {{A Unifying Framework}} to {{Benchmark NeuroAI Models}}},
  shorttitle = {{{NeuralBench}}},
  author = {Banville, Hubert and {d'Ascoli}, St{\'e}phane and Dahan, Simon and Rapin, J{\'e}r{\'e}my and Careil, Marl{\`e}ne and Benchetrit, Yohann and L{\'e}vy, Jarod and Panchavati, Saarang and Ratouchniak, Antoine and Mingfang and Zhang and Cascardi, Elisa and Begany, Katelyn and Brooks, Teon and King, Jean-R{\'e}mi},
  year = 2026,
  month = may,
  number = {arXiv:2605.08495},
  eprint = {2605.08495},
  primaryclass = {cs},
  publisher = {arXiv},
  doi = {10.48550/arXiv.2605.08495},
  archiveprefix = {arXiv}
}

@misc{xiongEEGFMBenchComprehensiveBenchmark2025,
  title = {{{EEG-FM-Bench}}: {{A Comprehensive Benchmark}} for the {{Systematic Evaluation}} of {{EEG Foundation Models}}},
  shorttitle = {{{EEG-FM-Bench}}},
  author = {Xiong, Wei and Li, Jiangtong and Li, Jie and Zhu, Kun and Jiang, Changjun},
  year = 2025,
  publisher = {arXiv},
  doi = {10.48550/ARXIV.2508.17742},
  copyright = {Creative Commons Attribution 4.0 International}
}

@article{delorme2004eeglab,
  title={{EEGLAB}: an open source toolbox for analysis of single-trial {EEG} dynamics including independent component analysis},
  author={Delorme, Arnaud and Makeig, Scott},
  journal={Journal of Neuroscience Methods},
  volume={134},
  number={1},
  pages={9--21},
  year={2004},
  doi={10.1016/j.jneumeth.2003.10.009}
}

@article{bendr2021,
  title={{BENDR: using transformers and a contrastive self-supervised learning task to learn from massive amounts of {EEG} data}},
  author={Kostas, Demetres and Aroca-Ouellette, Stephane and Rudzicz, Frank},
  journal={Frontiers in Human Neuroscience},
  volume={15},
  pages={653659},
  year={2021},
  doi={10.3389/fnhum.2021.653659}
}

@article{kuruppuEEGFoundationModels2026,
  title = {{{EEG}} Foundation Models: A Critical Review of Current Progress and Future Directions},
  shorttitle = {{{EEG}} Foundation Models},
  author = {Kuruppu, Gayal and Wagh, Neeraj and Kremen, Vaclav and Varatharajah, Yogatheesan},
  year = 2026,
  month = mar,
  journal = {Journal of Neural Engineering},
  volume = {23},
  number = {2},
  pages = {021001},
  publisher = {IOP Publishing},
  issn = {1741-2552},
  doi = {10.1088/1741-2552/ae4455},
  langid = {english}
}

@article{nagarajan2024healthy_to_stroke,
  author  = {Nagarajan, Aarthy and Robinson, Neethu and Ang, Kai Keng
             and Chua, Karen Sui Geok and Chew, Effie and Guan, Cuntai},
  title   = {Transferring a Deep Learning Model from Healthy Subjects
             to Stroke Patients in a Motor Imagery
             Brain--Computer Interface},
  journal = {Journal of Neural Engineering},
  year    = {2024},
  month   = jan,
  volume  = {21},
  number  = {1},
  pages   = {016007},
  doi     = {10.1088/1741-2552/ad152f}
}

@article{park2016eeg_lesion,
  author  = {Park, Wanjoo and Kwon, Gyu Hyun and Kim, Yun-Hee
             and Lee, Jong-Hwan and Kim, Laehyun},
  title   = {{EEG} Response Varies with Lesion Location in Patients
             with Chronic Stroke},
  journal = {Journal of NeuroEngineering and Rehabilitation},
  year    = {2016},
  month   = mar,
  volume  = {13},
  pages   = {21},
  doi     = {10.1186/s12984-016-0120-2}
}

@article{ang2011large_clinical,
  author  = {Ang, Kai Keng and Guan, Cuntai and Chua, Karen Sui Geok
             and Ang, Beng Ti and Kuah, Christopher Wee Keong
             and Wang, Chuanchu and Phua, Kok Soon
             and Chin, Zheng Yang and Zhang, Haihong},
  title   = {A Large Clinical Study on the Ability of Stroke Patients
             to Use an {EEG}-Based Motor Imagery
             Brain--Computer Interface},
  journal = {Clinical EEG and Neuroscience},
  year    = {2011},
  month   = oct,
  volume  = {42},
  number  = {4},
  pages   = {253--258},
  doi     = {10.1177/155005941104200411}
}

@article{shu2018bci_inefficient,
  author  = {Shu, Xiaokang and Chen, Shugeng and Yao, Lin
             and Sheng, Xinjun and Zhang, Dingguo and Jiang, Ning
             and Jia, Jie and Zhu, Xiangyang},
  title   = {Fast Recognition of {BCI}-Inefficient Users Using
             Physiological Features from {EEG} Signals:
             A Screening Study of Stroke Patients},
  journal = {Frontiers in Neuroscience},
  year    = {2018},
  month   = feb,
  volume  = {12},
  pages   = {93},
  doi     = {10.3389/fnins.2018.00093}
}

@article{delorme2023eeg,
  author  = {Delorme, Arnaud},
  title   = {{EEG} is better left alone},
  journal = {Scientific Reports},
  year    = {2023},
  volume  = {13},
  pages   = {2372},
  doi     = {10.1038/s41598-023-27528-0}
}

@article{braun2017mi_impairment,
  author  = {Braun, Niclas and Kranczioch, Cornelia and Liepert, Joachim
             and Dettmers, Christian and Zich, Catharina and B{\"u}sching, Imke
             and Debener, Stefan},
  title   = {Motor Imagery Impairment in Postacute Stroke Patients},
  journal = {Neural Plasticity},
  volume  = {2017},
  pages   = {4653256},
  year    = {2017},
  doi     = {10.1155/2017/4653256}
}

@article{kaiser2012stroke_erd,
  author  = {Kaiser, Vera and Daly, Ian and Pichiorri, Floriana
             and Mattia, Donatella and M{\"u}ller-Putz, Gernot R.
             and Neuper, Christa},
  title   = {Relationship Between Electrical Brain Responses to Motor Imagery
             and Motor Impairment in Stroke},
  journal = {Stroke},
  volume  = {43},
  number  = {10},
  pages   = {2735--2740},
  year    = {2012},
  doi     = {10.1161/STROKEAHA.112.665489}
}

@article{pfurtscheller2006mu_classification,
  author  = {Pfurtscheller, Gert and Brunner, Clemens and Schl{\"o}gl, Alois
             and Lopes da Silva, Fernando H.},
  title   = {Mu Rhythm (De)Synchronization and {EEG} Single-Trial
             Classification of Different Motor Imagery Tasks},
  journal = {NeuroImage},
  volume  = {31},
  number  = {1},
  pages   = {153--159},
  year    = {2006},
  doi     = {10.1016/j.neuroimage.2005.12.003}
}

@article{chen2021erd_accuracy,
  author  = {Chen, Shugeng and Shu, Xiaokang and Wang, Hewei and Ding, Li
             and Fu, Jianghong and Jia, Jie},
  title   = {The Differences Between Motor Attempt and Motor Imagery in
             Brain-Computer Interface Accuracy and Event-Related
             Desynchronization of Patients With Hemiplegia},
  journal = {Frontiers in Neurorobotics},
  volume  = {15},
  pages   = {706630},
  year    = {2021},
  doi     = {10.3389/fnbot.2021.706630}
}

@inproceedings{kwon2018erd_performance,
  author    = {Kwon, Moonyoung and Cho, Hohyun and Won, Kyungho
               and Ahn, Minkyu and Jun, Sung Chan},
  title     = {Event-Related Desynchronization ({ERD}) May Not Be Correlated
               With Motor Imagery {BCI} Performance},
  booktitle = {2018 IEEE International Conference on Systems, Man, and
               Cybernetics (SMC)},
  pages     = {1133--1137},
  year      = {2018},
  doi       = {10.1109/SMC.2018.00200}
}



\end{document}